\documentclass[a4paper,11pt,final]{article}
\usepackage[utf8]{inputenc}
\usepackage[T1]{fontenc}
\usepackage[english]{babel}
\usepackage{amsmath}
\usepackage{amssymb}
\usepackage{amsthm}
\usepackage{dsfont}  
\usepackage{fancyhdr}  
\usepackage{color}
\usepackage{graphicx}
\usepackage{a4wide}
\usepackage{eucal}  

\usepackage{showkeys}  

\graphicspath{{./Images/}{./}}

\newcommand{\ssi}{\Longleftrightarrow}
\newcommand{\prive}{\backslash}

\makeatletter
\newcommand{\dfn}{\@ifstar\dfnStar\dfnNostar}
\newcommand{\dfnStar}[1]{\emph{#1}}
\newcommand{\dfnNostar}[1]{{\bf \emph{#1}}}
\makeatother

\newcommand{\A}{\mathcal{A}}
\newcommand{\B}{\mathcal{B}}

\newcommand{\D}{\mathcal{D}}
\newcommand{\F}[1][]{\mathcal{F}_{#1}}
\newcommand{\G}{\mathcal{G}}
\newcommand{\I}{\mathcal{I}}
\renewcommand{\P}{\mathcal{P}}
\renewcommand{\S}{\mathcal{S}}
\newcommand{\NN}{\mathbb{N}}
\newcommand{\RR}{\mathbb{R}}
\newcommand{\ZZ}{\mathbb{Z}}
\newcommand{\QQ}{\mathbb{Q}}

\newcommand{\pg}{{\bf p}}
\newcommand{\Sb}{\mathcal{T}_b}

\newcommand{\pc}[1]{{#1}_+^c}
\newcommand{\ms}[1]{{#1}_-^*}
\newcommand{\ps}[1]{{#1}_+^*}
\newcommand{\pequ}[1]{\stackrel{\neq {#1}}{=}}
\newcommand{\db}{\underline{\partial}}
\newcommand{\ol}[1]{\overline{#1}}

\newcommand{\join}[1][]{\stackrel{#1}{\rightsquigarrow}}
\newcommand{\dom}{\preccurlyeq}

\DeclareMathOperator{\indicatrice}{\mathds{1}}
\newcommand{\1}[1]{\indicatrice_{#1}}

\DeclareMathOperator{\Supp}{Supp}
\DeclareMathOperator{\e}{e}

\renewcommand{\leq}{\leqslant}
\renewcommand{\geq}{\geqslant}
\renewcommand{\emptyset}{\varnothing}
\renewcommand{\epsilon}{\varepsilon}
\renewcommand{\rho}{\varrho}
\renewcommand{\phi}{\varphi}

\numberwithin{equation}{section}
\numberwithin{figure}{section}

\theoremstyle{plain}
\newtheorem{THEOREME}{Theorem}[section]
\newtheorem{PROPOSITION}[THEOREME]{Proposition}
\newtheorem{COROLLAIRE}[THEOREME]{Corollary}
\newtheorem{LEMME}[THEOREME]{Lemma}

\theoremstyle{definition}
\newtheorem{DEFINITION}[THEOREME]{Definition}
\newtheorem{REMARQUE}[THEOREME]{Remark}
\newtheorem{REMARQUES}[THEOREME]{Remarks}
\newtheorem{EXEMPLE}[THEOREME]{Example}
\newtheorem{EXEMPLES}[THEOREME]{Examples}

\newenvironment{theoreme}[1][]
    {\begin{THEOREME}[#1]\ \par }
    {\end{THEOREME}}
\newenvironment{proposition}[1][]
    {\begin{PROPOSITION}[#1]\ \par }
    {\end{PROPOSITION}}
\newenvironment{corollaire}[1][]
    {\begin{COROLLAIRE}[#1]\ \par }
    {\end{COROLLAIRE}}
\newenvironment{lemme}[1][]
    {\begin{LEMME}[#1]\ \par }
    {\end{LEMME}}
\newenvironment{preuve}[1][Proof]
    {\begin{proof}[#1]\ \par}
    {\end{proof}}
\newenvironment{definition}[1][]
    {\begin{DEFINITION}[#1]\ \par }
    {\end{DEFINITION}}
\newenvironment{remarque}
    {\begin{REMARQUE} \ \par }
    {\end{REMARQUE}}
\newenvironment{remarques}
    {\begin{REMARQUES} \ \par }
    {\end{REMARQUES}}
\newenvironment{exemple}[1][]
    {\begin{EXEMPLE}[#1]\ \par }
    {\end{EXEMPLE}}
\newenvironment{exemples}
    {\begin{EXEMPLES}\ \par }
    {\end{EXEMPLES}}
    
\newcommand{\reff}[1]{(\ref{#1})}
\newcommand{\bydef}{:=}

\title{Partially ordered models}
\author{Vincent Deveaux and Roberto Fern\'andez\footnote{Present address: 
    Department of Mathematics,
    Utrecht University,
    P.O. Box 80010 
    3508 TA Utrecht, The Netherlands, \tt{R.Fernandez1@uu.nl}
  }\\
  \small{Laboratoire de Math\'ematiques Rapha{\"e}l Salem}\\
  \small{UMR 6085 CNRS-Universit\'e de Rouen}\\
  \small{Avenue de l'Universit\'e, BP.12}\\
  \small{76801 Saint Etienne du Rouvray - France}\\
  \small{\tt{vincent.deveaux@univ-rouen.fr}}\\
  \small{\tt{roberto.fernandez@univ-rouen.fr}}
}

\date{February 20, 2010\\
submited to Journal of Statistical Physics, February 22, 2010}

\begin{document}

\maketitle

\begin{abstract}
  We provide a formal definition and study the basic properties of partially ordered chains (POC).
  These systems were proposed to model textures in image processing and to represent independence relations between random variables in statistics (in the later case they are known as Bayesian networks).
  Our chains are a generalization of probabilistic cellular automata (PCA) and their theory has features intermediate between that of discrete-time processes and the theory of statistical mechanical lattice fields.
  Its proper definition is based on the notion of partially ordered specification (POS), in close analogy to the theory of Gibbs measure.
  This paper contains two types of results.
  First, we present the basic elements of the general theory of POCs:
  basic geometrical issues, definition in terms of conditional probability kernels, extremal decomposition, extremality and triviality, reconstruction starting from single-site kernels, relations between POM and Gibbs fields.
  Second, we prove three uniqueness criteria that correspond to the criteria known as bounded uniformity, Dobrushin and disagreement percolation in the theory of Gibbs measures. 
\end{abstract}


\tableofcontents

%
%
\section{Framework}
\label{todo:intro-part-1}

The name \emph{partially ordered Markov model} (POMM) first appeared in two articles in statistics \cite{cressie-davidson,Davidson-Cressie-Hua} dealing with the analysis of black-and-white textures in images.
The authors described some basic features of the models and showed its efficiency for the storing and simulation of some well-chosen textures.
Independently, closely related models ---called \emph{Bayesian networks}--- have been used, also in statistics, to model networks of conditional independence relations between large numbers of random variables (see, for instance, \cite{jennie07}).  
These networks need not be Markovian, in fact the Markovian version is also known as \emph{Markov blankets}.
For concreteness, we call \emph{partially ordered model} (POM) the general, non-necessarily Markovian, version which is the object of our work.  

The increasing popularity of these networks justifies, in our opinion, their formal study as probabilistic objects.
Indeed, these objects have a number of interesting features which place them in between two vastly studied categories of models ---probabilistic cellular automata (PCA) and lattice Gibbsian fields.
On the one hand, POMs generalize PCAs by replacing the totally ordered time axis by a partially ordered lattice.
On the other hand, POMs are also random fields described by finite-region conditional probabilities which, however, are measurable only with respect to the partial past, rather than to the whole exterior of the region as in the Gibbsian case.  

In this paper, we first discuss the proper definition of POMs as measures consistent with appropriate conditional kernels.
Some geometrical issues need to be settled, regarding allowed regions ---\emph{good regions} or \emph{time boxes}--- for the development of the theory.
These are regions whose (partial) past is separated from the (partial) future, a fact that prevents measurability conflicts.
We also explicitly determine the ``(re)construction'' procedure that yields the whole of the specification starting from single-site kernels.
The existence of this procedure justifies the study of POMs only in terms of single-site conditional probabilities, as it is usually done, without warning, in the literature.
This is in analogy to the study of PCAs, which is based in single-time transition kernels.
Next, we turn to the general ``phase diagram'' theory, developed along the lines of Gibbsian theory.
In general, a partially ordered specification can have several consistent measures, each of which we call a \emph{partially ordered chain} (POC).
We show that chains that are extremal under convex decompositions satisfy tail-field triviality and mixing properties analogous to those of phases in statistical mechanics.
To further the study of phase diagrams, we also discuss FKG-like inequalities for POMs.

The present ``statistical mechanical'' treatment generalizes work done for discrete-time processes \cite{fernandez-maillard}.
Related issues were addressed for PCAs in the fundamental work done in \cite{lebowitz-maes-speer-2,lebowitz-maes-speer}; see also \cite{pca2gibbsFields}.
In these references, the statistical mechanical features of the theory of PCAs are studied by relating them to Gibbs fields in one more dimension.
Unlike the present paper, this strategy involves a restriction to translation-invariant PCAs.

The second family of results presented here involves a series of \emph{uniqueness criteria}, that is, conditions under which a POM admits only one consistent POC.
We present three different criteria that correspond to similar results within the theory of Gibbsian measures:
\begin{itemize}
\item[(i)] \emph{Bounded uniformity criterion:}
  There is a unique consistent POC if the effect of changing boundary conditions is bounded by multiplicative factors at the level of kernels.
  In a Gibbsian setting, this corresponds to finite energy differences between external conditions.
  Such a condition explains, for instance, why all Markov ---or, more generally, finite-range or tail-summable--- one-dimensional models do not exhibit phase transitions.
  In our setting, the criterion is also useful only for models that are, in some sense, ``one-dimensional''. 
\item[(ii)] \emph{Dobrushin criterion:}
  A POM has a unique POC if the sum of the oscillations of the single-site kernels is smaller than one.
  This sum of oscillations is, in most cases, numerically computable, a fact that opens the way for computer-assisted proofs~\cite{Kalafa}.
  Such a criterion generalizes the Dobrushin criterion previously proven both for Markovian PCAs~\cite{maes-shlosman} and for (non-necessarily Markovian) chains~\cite{fernandez-maillard-2}.
\item[(iii)] \emph{Disagreement percolation criterion:}
  A duplicated system is proposed and the sites where both copies disagree are registered.
  Uniqueness holds if a coupling ---that is, a simultaneous realization of both copies--- can be defined such that these disagreement sites do not percolate.
  This criterion, which has been very successful for Gibbsian measures \cite{vandenberg,VanDenBerg-Maes}, applies only in the Markovian framework.
\end{itemize}

Through our paper, we illustrate our results through two simple but revealing  examples: the POMM-Ising and Stavskaya models.

%
%
\section{Set-up and examples}

%
\subsection{The issue}

The basic ingredients of our models are: 
\begin{itemize}
\item[(i)]
  A countable (partially) ordered set $(S,\leq)$, called the \dfn{space of sites}.
  Each site $x\in S$ determines a \dfn{past} $x_-=\bigl\{y\in S: y<x\bigr\}$ and a \dfn{future} $x_+=\bigl\{y\in S: y>x\bigr\}$ .
\item[(ii)]
  A measurable set $(E,\mathcal{E})$, the \dfn{space of colors}, which, in general, needs not be supposed either finite or countable.
\item[(iii)]
  The product space $(\Omega, \F)=(E^S, \mathcal{E}^S)$ ---the \dfn{configuration space}.
  If $\Upsilon\subset S$, $\F[\Upsilon]$ denotes the sub-$\sigma$-algebra of $\F$ generated by the cylinders with base in $E^\Upsilon$.
  We shall use lowercase Greek letters ---$\omega$, $\sigma$, \ldots--- for configurations in $\Omega$, and the restriction of a configuration $\omega$ to a set of sites $\Upsilon$ will be denoted $\omega_\Upsilon$.
  If $\Upsilon=\{x\}$ the braces will be omitted ($\omega_x, \F[x]$, etc).
\item[(iv)]
  A family of \dfn{single-site oriented kernels} $\bigl\{\gamma_x(\,\cdot\,,\,\cdot\,): x\in S\bigr\}$, where each $\gamma_x(d\xi,\omega)$ is a probability measure on $\F[\pc{x}]$ with respect to the first argument and a measurable function with respect to the second one.
  The kernel is oriented in that for every event $A$ on the site $x$, $\gamma_x(A.\,\cdot\,)$ depends only on the past $x_-$. 
\end{itemize}

The object of our study are measures $\mu$ on $(\Omega,\F)$ that are \emph{consistent} with the kernels $\{\gamma_x\}$.
Consistency can be understood in two senses:
\begin{itemize}
\item[(C1)]
  $\mu$ is the limit of the iterated application of the kernels from past to future.  
\item[(C2)]
  $\mu$ is such that $\gamma_x$ is its conditional expectations on the site $x$ given its past.
\end{itemize}

In the sections that follow we present the proper mathematical statements of these two characterizations and the proof of their equivalence.
Before that, let us turn to two examples to be used as illustration in the rest of the paper.

%
\subsection{Benchmark examples}
\label{ss:benchmark}

Both our examples refers to $S=\ZZ^2$ with the natural partial order
\begin{equation}
  \label{eqn:ZZ2}
  (x_1,y_1)\leq (x_2,y_2)\ \ssi\  x_1\leq x_2 \text{ and } y_1\leq y_2\;.
\end{equation}

Unless otherwise specified, this will be the order understood in the drawings of our article.
Furthermore, in these drawings the vertical axis will be oriented downwards, following a widespread convention in computer science.   

Our benchmark models are defined in terms of the spatial operators:
\[\begin{split}
  N : \ZZ^2\to \ZZ^2 &: x=(x_1,x_2) \mapsto Nx=(x_1  ,x_2-1) \\
  W : \ZZ^2\to \ZZ^2 &: x=(x_1,x_2) \mapsto Wx=(x_1-1  ,x_2) \\
\end{split}\]
[$N$ stands for Northern neighbor and $W$ for Western neighbor].

%
\subsubsection{The POMM-Ising Model}

This model is the partially oriented version of the 2-dimensional Ising model in statistical mechanics.
The color space has only two elements: $E=\{-1,+1\}$ and the kernels are determined by two parameters $h\in\RR$ and $\beta>0$.
By analogy to its statistical-mechanical meaning, we shall call $h$ the magnetic field and $\beta$ the inverse temperature. The single-site kernels are of the form:
\begin{equation}
  \label{eqn:POMM-Ising}
  \forall\xi\in\Omega,\ \forall\sigma\in E,\quad
  \I_{\{x\}}(\sigma,\xi) := \frac{1}{Z_\xi}\exp\Bigl[\beta\sigma\bigl(\xi_{Nx}+\xi_{Wx}+h\bigr)\Bigr]
\end{equation}
where $Z_\xi$ is the normalizing coefficient:
\[
Z_\xi := \exp\Bigl[-\beta(\xi_{Nx}+\xi_{Wx}+h)\Bigr] + \exp\Bigl[\beta(\xi_{Nx}+\xi_{Wx}+h)\Bigr]
\]

Note that if $h=0$ the model becomes a voter model:
\begin{equation}
  \label{eqn:voter model}
  \I_{\{x\}}(\sigma,\xi) =
  \begin{cases}
    1-\epsilon & \text{if }\xi_{Nx}=\xi_{Wx}=\sigma \\
    1/2 & \text{if }\xi_{Nx}\neq\xi_{Wx} \\
  \end{cases}
\end{equation}
where $\epsilon = \e^{-2\beta} /\bigl[\e^{2\beta}+\e^{-2\beta}\bigr] \;\in (0,1/2)$.

\begin{figure}[htbp]
  \begin{center}
    \begin{tabular}{ccc}
      \includegraphics[width=4.5cm]{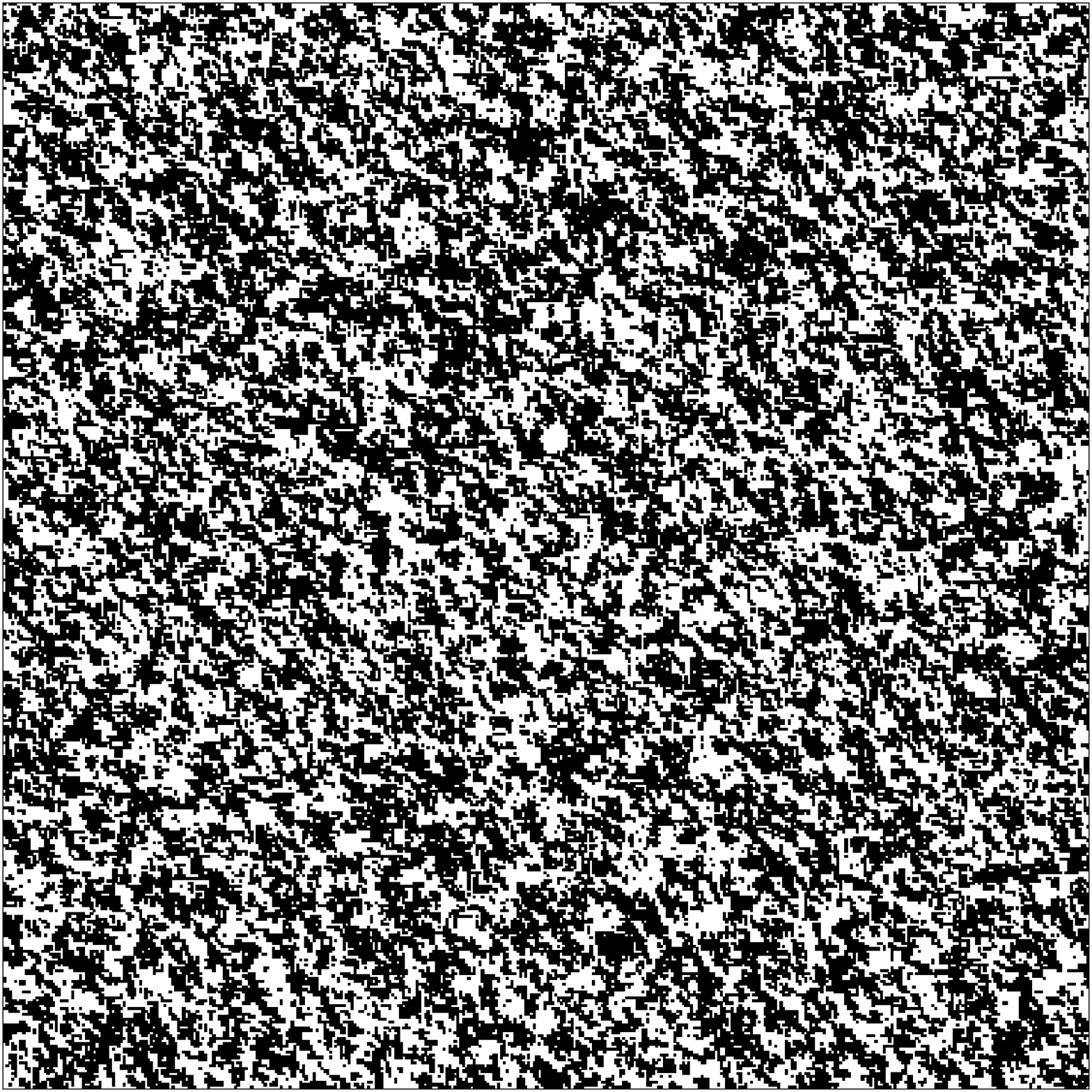} & \includegraphics[width=4.5cm]{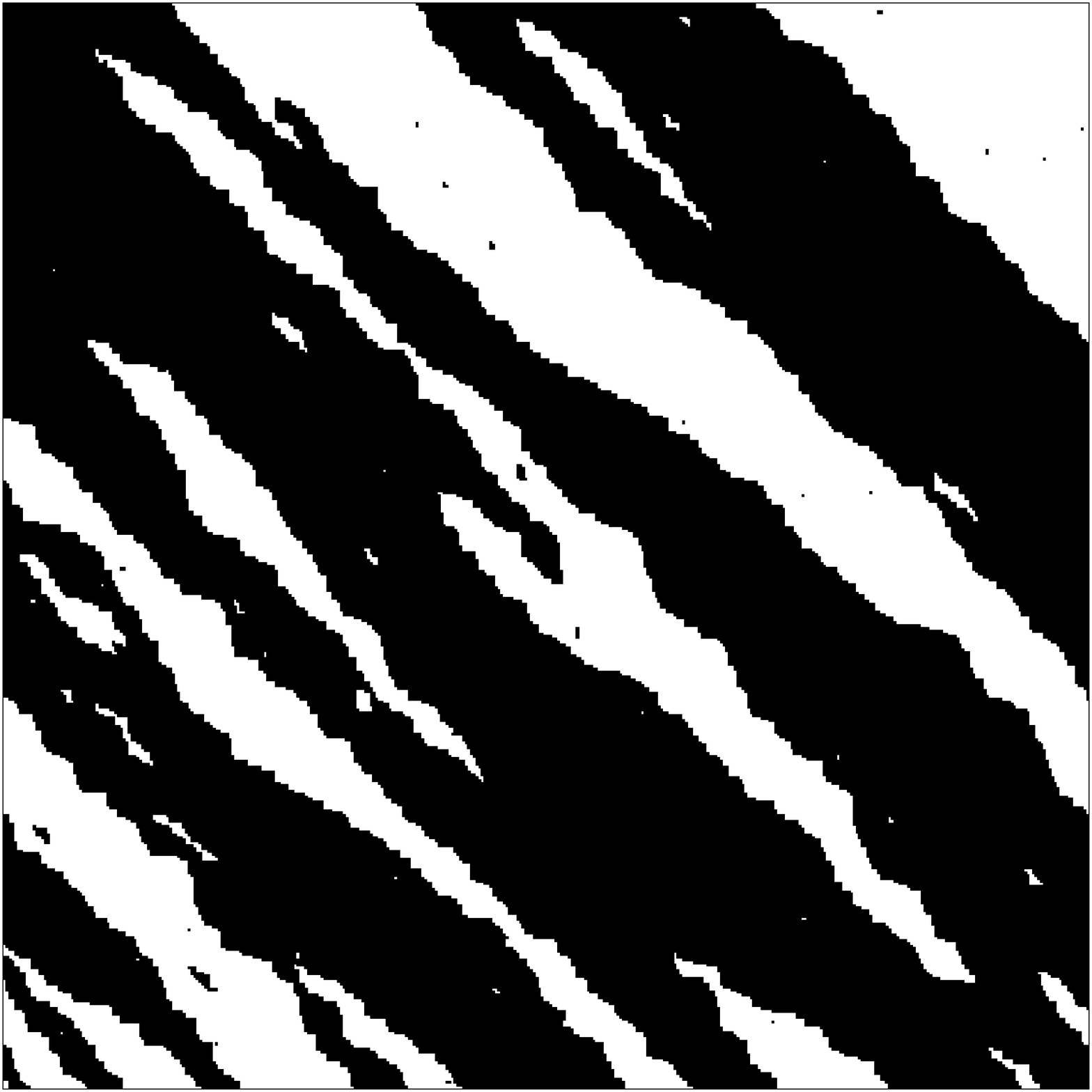} & \includegraphics[width=4.5cm]{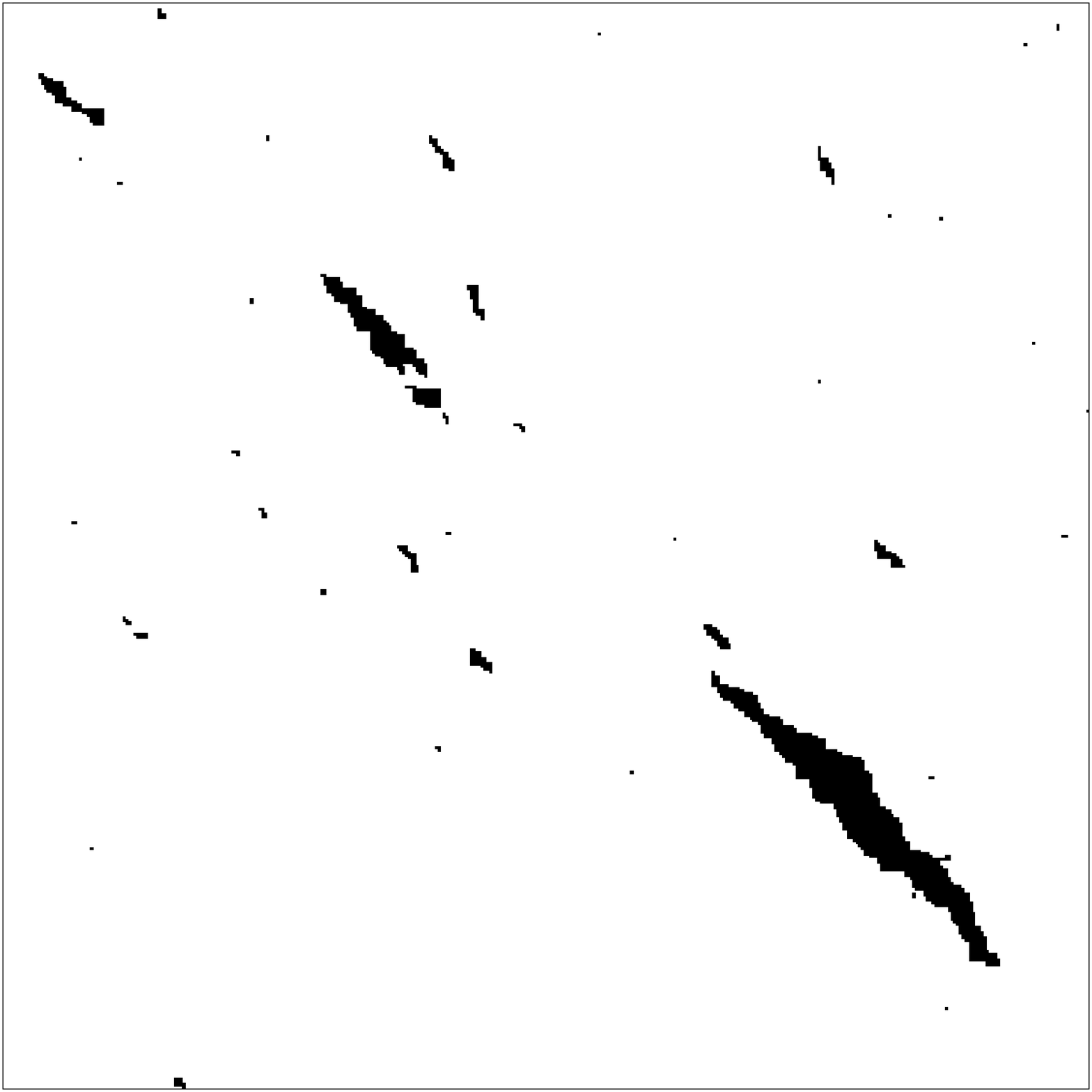} \\
      $(\beta,h)=(0.5,\ 0)$ & $(\beta,h)=(2,\ 0)$ & $(\beta,h)=(2,\ 0.05)$ \\
    \end{tabular}
    \caption{Simulations of the Pomm-Ising model.
      Color $1$ is black and $-1$ is white}
    \label{fig:simul pomm-ising}
  \end{center}
\end{figure}

Figure \ref{fig:simul pomm-ising} shows simulations of the resulting POC.
Without magnetic field, the low-$\beta$ (high temperature) configurations are very disordered while texture appears at high-$\beta$.
The behavior, however, is drastically changed by the presence of even a small magnetic field.

We show in \cite{pomm-geom} that this model has no phase transition:
for any value of $\beta$ and $h$ there exists only one $\I$-POC.

%
\subsubsection{The Stavskaya's Model}
\label{subsec:Stavskaya}

The color space is $E=\{0,1\}$, and the single site kernels are
\begin{equation}
  \label{eqn:or. perc.}
  \S_{\{x\}}(\sigma=1,\xi) :=
  \begin{cases}
    p & \text{if }\xi_{Nx}+\xi_{Wx}>0, \\
    0 & \text{otherwise}. \\
  \end{cases}
\end{equation}
for some $p\in[0,1]$.  

\begin{figure}[htbp]
  \begin{center}
    \begin{tabular}{ccc}
      \includegraphics[width=4.5cm]{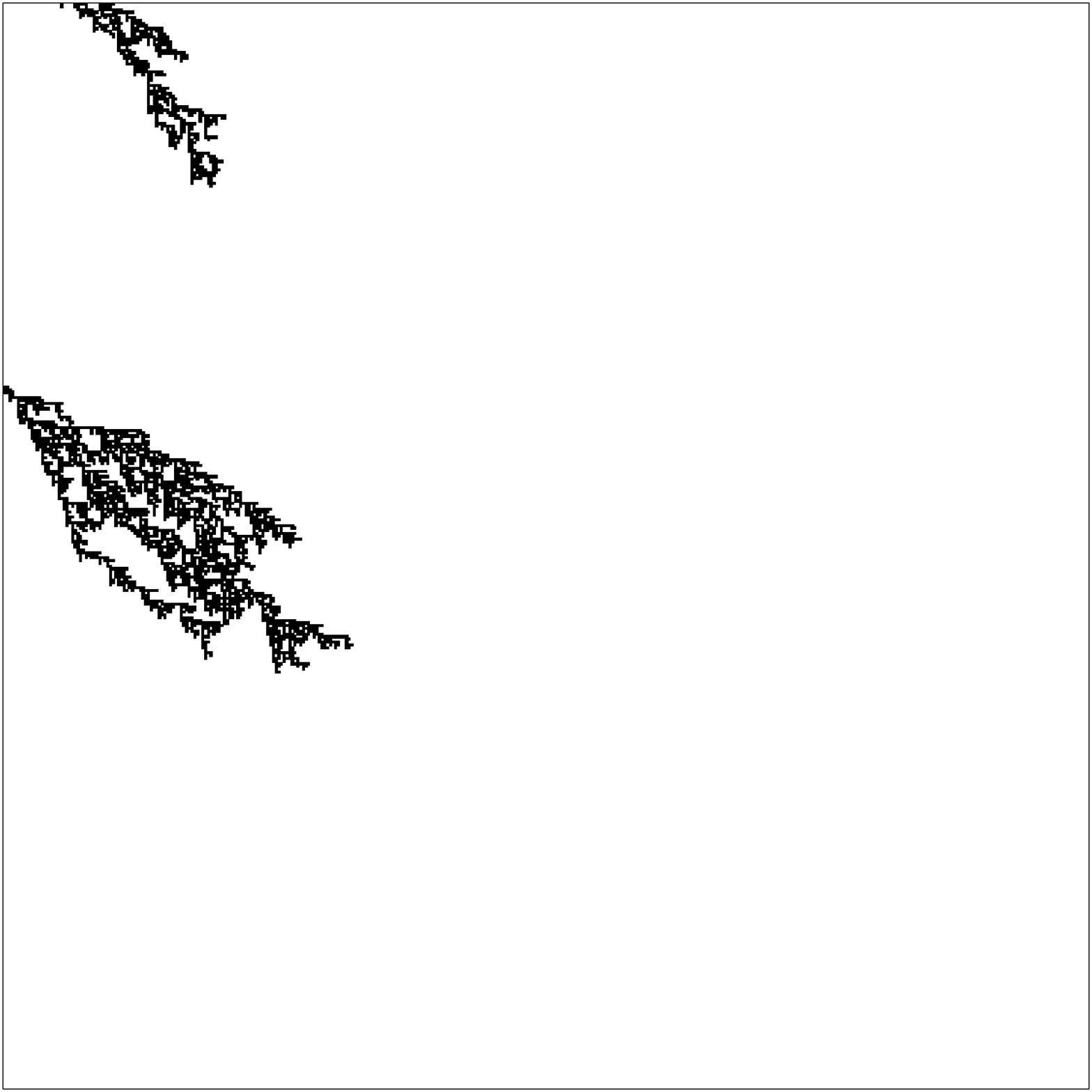} & \includegraphics[width=4.5cm]{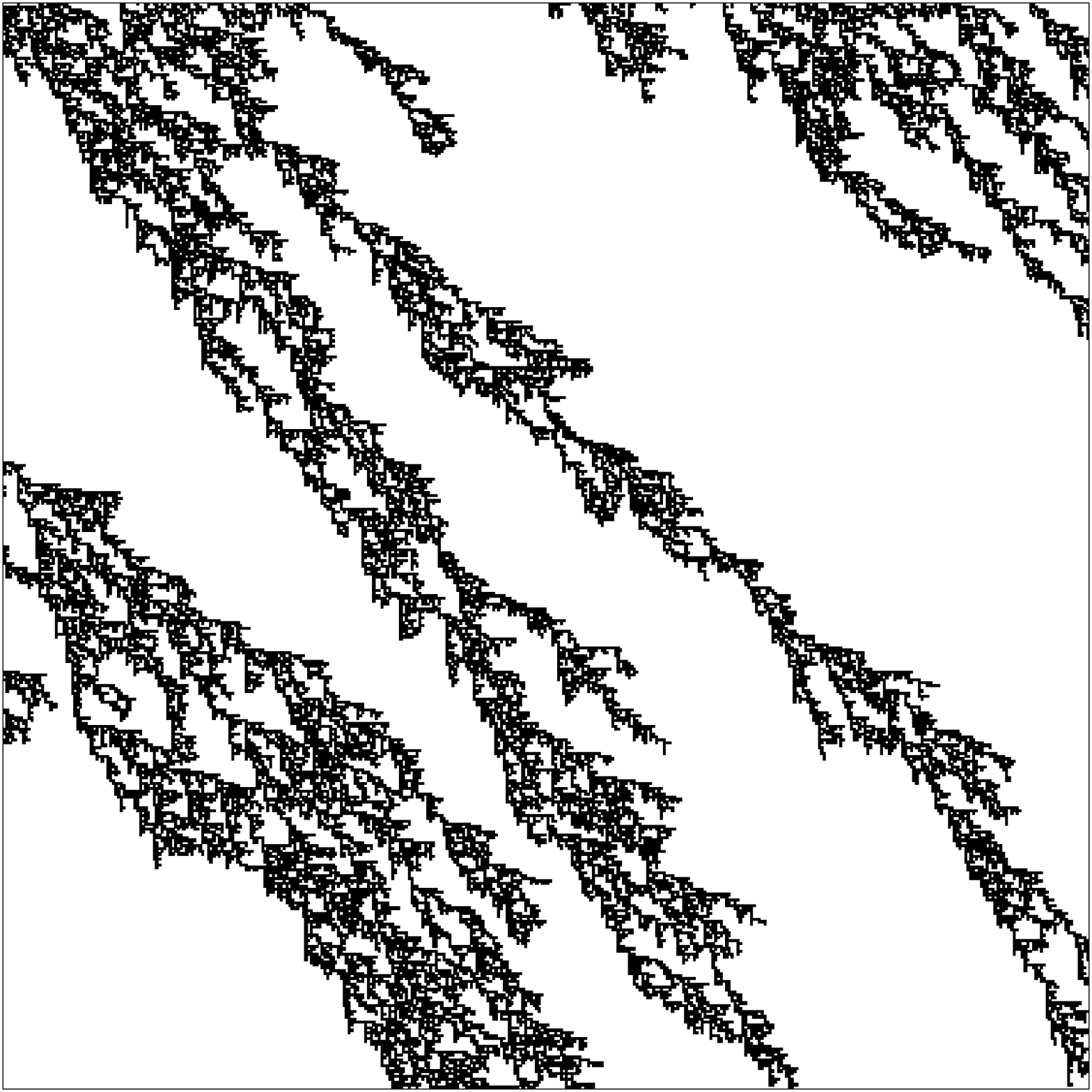} & \includegraphics[width=4.5cm]{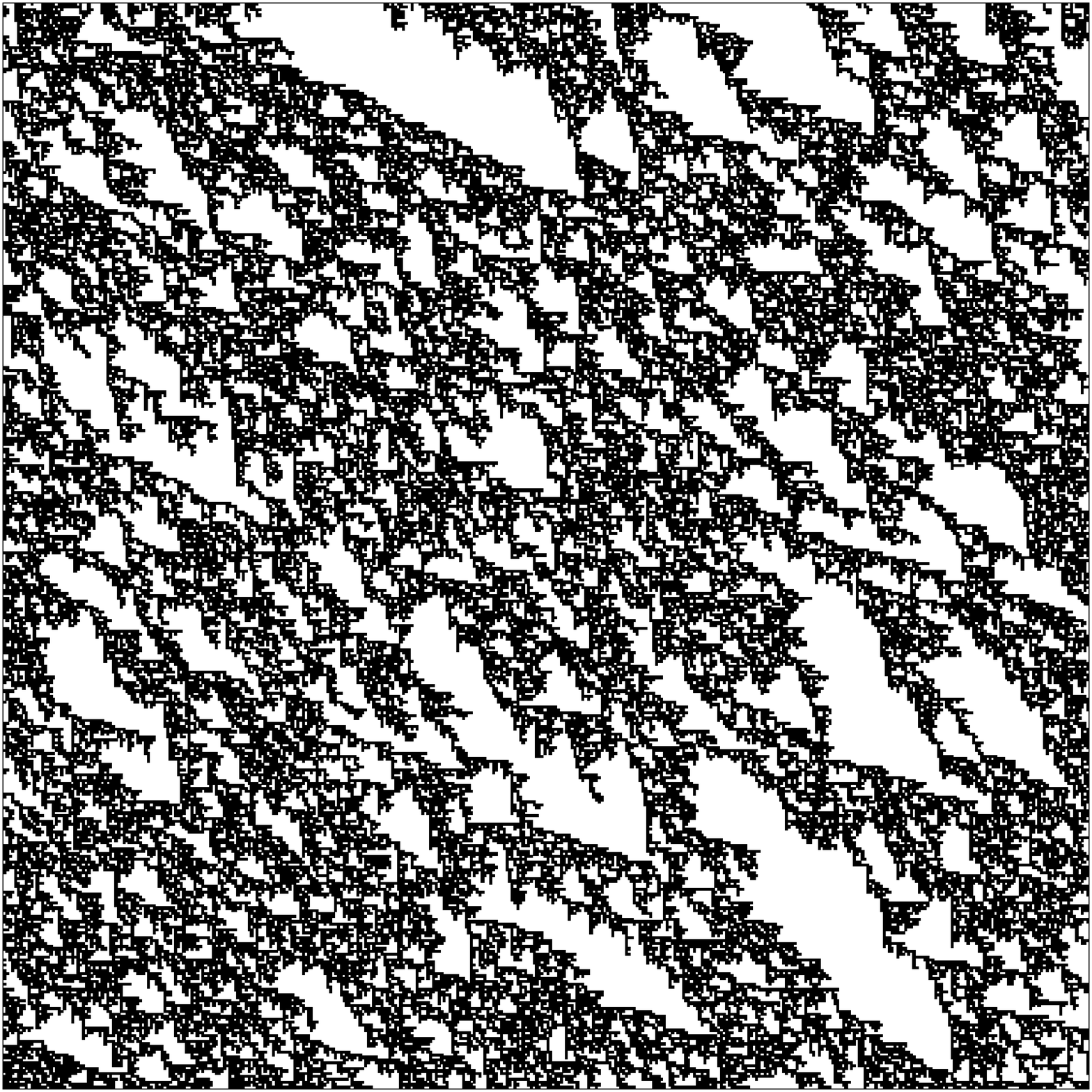} \\
      $p=0.7$ & $p=0.702$ & $p=0.72$ \\
    \end{tabular}
    \caption{Simulations of the Stavskaya's model.
      Color $0$ is white and $1$ is black.}
    \label{fig:simul stavskaya}
  \end{center}
\end{figure}

This model can be seen as a model for oriented percolation.
A site can become occupied ($\sigma_x=1$) only if one of its N or W neighbors is occupied.
Starting from the boundary of a box, we can see that this process produces clusters with the Bernoulli law of classical oriented percolation.
Existence of an infinite cluster in the latter is, therefore, equivalent to the existence of a POC $\nu$ with $\nu\bigl(\{\sigma_0=1\}\bigr)>0$.
Since the $\delta$-measure concentrated on the ``all $0$''-configuration is clearly always consistent with this POM, the existence of an infinite cluster in oriented independent percolation becomes equivalent to the existence of more than one POC, that is, on the occurrence of a phase transition.
The critical value $p_c^+$ for oriented percolation corresponds, then, to a value such that for $p<p_c^+$ there exists a unique $\S$-POC while uniqueness is lost for $p>p_c^+$.
The simulations in Figure \ref{fig:simul stavskaya} clearly show this transition.
For small $p$, the image is almost all white.
Some filaments of occupied sites began to appear at $p\sim 0.7$ and they built a thick network for larger $p$.  
The transition appears to be sharp.

%
%
\section{Formal definitions}
\label{sec:preliminaries-1}

%
\subsection{Geometrical aspects}

The proper definition of POMs requires some preliminary geometric considerations.
Let $(S,\leq)$ be a countable (partially) ordered set.

\begin{definition}
  Let $\Upsilon\subset S$.
  We define:
  \begin{itemize}
  \item[(i)]
    The \dfn{maxima} and \dfn{minima} of $\Upsilon$,
    \begin{equation}
      \label{eqn:minmax}
      \begin{split}
        \max(\Upsilon) & := \left\{ x\in\Upsilon;\ \forall y>x,\ y\not\in\Upsilon \right\} \\
        \min(\Upsilon) & := \left\{ x\in\Upsilon;\ \forall y<x,\ y\not\in\Upsilon\right\}\;. \\
      \end{split}
    \end{equation}
    
  \item[(ii)]
    The \dfn{past} of $\Upsilon$, 
    \begin{equation}
      \label{eqn:passé}
      \Upsilon_- := \left\{ x\in S, x\notin\Upsilon; \exists y\in\Upsilon, x<y \right\}\;.
    \end{equation}
    
  \item[(iii)]
    The \dfn{future} of $\Upsilon$,
    \begin{equation}
      \label{eqn:futur}
      \Upsilon_+ := \left\{ x\in S, x\notin\Upsilon; \exists y\in\Upsilon, x>y \right\}\;.
    \end{equation}
    
  \item[(iv)]
    The \dfn{outer time} of $\Upsilon$,
    \begin{equation}
      \label{eqn:etoile}
      \Upsilon^* := \left\{ x\in S; \forall y\in\Upsilon, x\text{ is unrelated to }y \right\}\;.
    \end{equation}
  \end{itemize}
\end{definition}

In the whole article we assume that the following properties hold for all $x\in S$:
\begin{itemize}
\item[(a)]
  $\max(\{x\}_-)$ is finite,
\item[(b)]
  $\min(\{x\}_+)$ is finite,
\item[(c)]
  $\forall y<x,\ \exists y_0\in\max(\{x\}_-),\ y\leq y_0 < x$, and
\item[(d)]
  $\forall z>x,\ \exists z_0\in\min(\{x\}_+),\ z\geq z_0 > x$.
\end{itemize}

Note that (c) and (d) imply that $\max(\{x\}_-)\neq\emptyset$ and $\min(\{x\}_+)\neq\emptyset$.
$S=\ZZ\times\QQ$ endowed with the ``natural'' partial order shows that this is not equivalent.

Moreover, we will suppose that $S$ does not contain minimal points: $\min(S)=\emptyset$.
This last hypothesis is not really crucial, but allows us to avoid the uninteresting case where there are sites at the infinite past.

\begin{definition}
  Let $\Lambda$ be a finite part of $S$.
  We say that $\Lambda$ is a \dfn{time box} if $\Lambda_-\cap\Lambda_+=\emptyset$, and a \dfn{bad box} otherwise.
  The set of time boxes is denoted by $\Sb$.
\end{definition}

Intuitively, time boxes have no holes because sites in a hole are in the past of some sites of the box but in the future of other box sites.
Our partially ordered kernels only make sense for time boxes, because they must depend on the past but, at the same time, are forbidden to say anything about the future.  
Let us see what a time box means in two typical cases.

\begin{figure}[htbp]
  \begin{center}
    \begin{tabular}{ccc}
      \includegraphics[width=3cm]{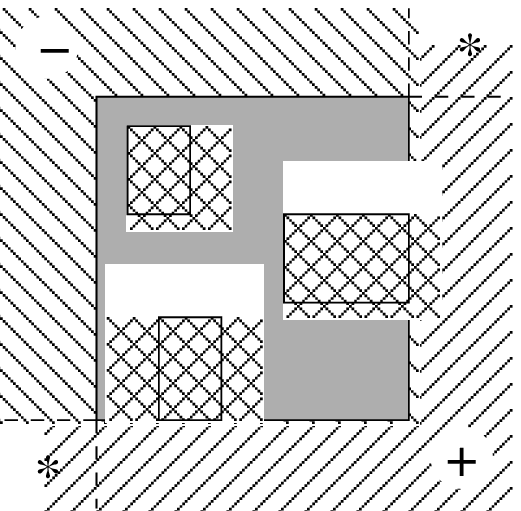} & \hspace{1cm} & \includegraphics[width=3cm]{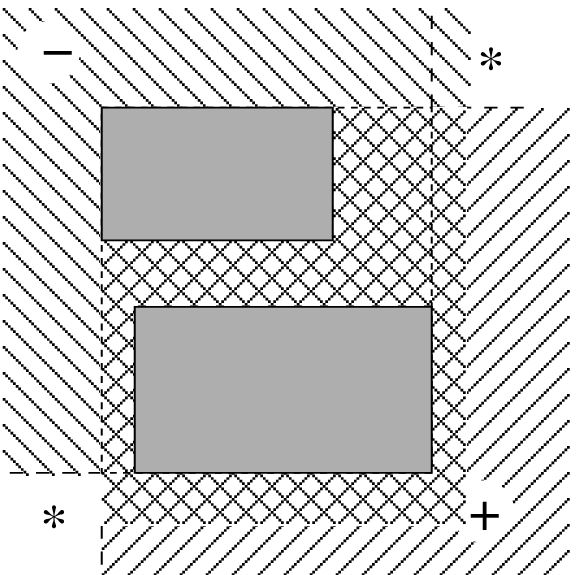} \\
      (a) & & (b) \\
      & & \\
      \includegraphics[width=3cm]{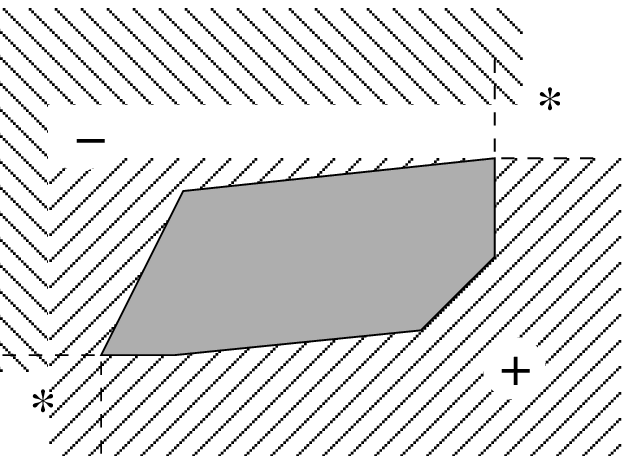} & \hspace{1cm} & \includegraphics[width=3cm]{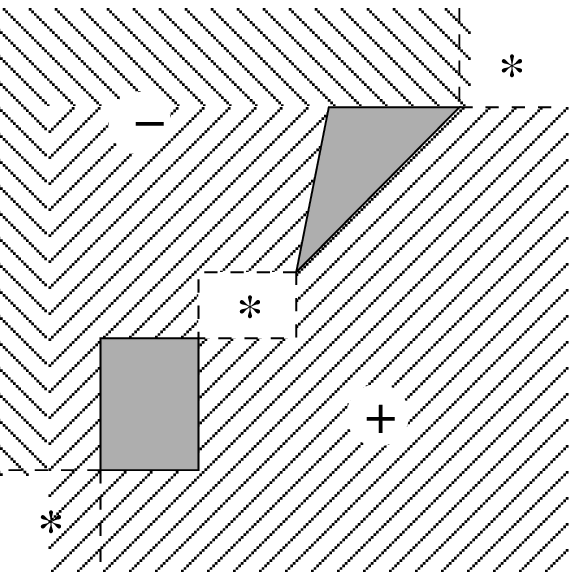} \\
      (c) & & (d) \\
    \end{tabular}
  \end{center}
  \caption{$\ZZ^2$ is endowed with its natural partial order \eqref{eqn:ZZ2}.
    The orientation is from the left-up corner to the right-down one, according to computer science conventions.
    If we denote $\Lambda$ the grey box, the symbol * stands for $\Lambda^*$, - for $\Lambda_-$ and + for $\Lambda_+$.
    (a), (b): bad boxes; (c), (d): time boxes.}
  \label{fig:boxes}
\end{figure}

\begin{exemples}\ \par
  \begin{enumerate} 
  \item
    $S=\ZZ$ with its natural (total) order:
    Since it is a total order, there is no outer time and $\Sb$ is exactly the set of finite intervals.
    The resulting POMs are, in fact, the left interval specifications \cite{fernandez-maillard,maillard}) found in the study of discrete-time processes.
  \item
    $S=\ZZ^2$ with the natural partial order \reff{eqn:ZZ2}.
    Figure \ref{fig:boxes} shows examples of time boxes and bad boxes.
  \end{enumerate}
\end{exemples}

In the sequel we will denote time boxes by $\Lambda$, $\Delta$ or $\Gamma$, and reserve $\Upsilon$ for subsets of $S$ without any assumption.
For a time box $\Lambda$, let us denote
\begin{equation}
  \pc{\Lambda} \;:=\; S\prive\Lambda_+\quad,\quad
  \ms{\Lambda} \;:=\; \Lambda_-\cup\Lambda^*\;.
\end{equation}
For $x\in S$, we shall abbreviate $x_- := \{x\}_-$ and $x_+ := \{x\}_+$.

\begin{remarques}
  \label{rmq:cap ms}\ \par
  \begin{enumerate}
  \item $\Sb$ is not empty.
    Indeed, for every single site $x\in S$, $\{x\}$ is a time box.
  \item For any time box $\Lambda$, $\{\Lambda,\ \Lambda_+,\ \Lambda_-,\ \Lambda^*\}$ is a partition of $S$.
  \item If $\Lambda$ is a time box, $\ms{\Lambda}=\pc{\Lambda}\prive\Lambda$ and $\Lambda_-\subset\ms{\Lambda}\subset\pc{\Lambda}$.
  \end{enumerate}
\end{remarques}

%
\subsection{Probabilistic notions}

We turn now to the definition of kernels (POMs) and consistent measures (POCs). 

\begin{definition}
  \label{def:POK}
  Let $\Lambda$ be a time box.
  A \dfn{proper oriented kernel} $\gamma_\Lambda$ on $\Lambda$ is a function $\gamma_\Lambda:\F[\pc{\Lambda}]\times\Omega\longrightarrow[0,1]$ satisfying the following properties:
  \begin{description}
  \item[(i)]
    For each $\omega\in\Omega$, $\gamma_\Lambda(\cdot,\omega)$ is a probability measure,
  \item[(ii)]
    For each $A\in\F[\pc{\Lambda}]$, $\gamma_\Lambda(A,\cdot)$ is $\F[\ms{\Lambda}]$-measurable,
  \item[(iii)]
    For each $A\in\F[\Lambda]$, $\gamma_\Lambda(A,\cdot)$ is $\F[\Lambda_-]$-measurable,
  \item[(iv)]
    For each $B\in\F[\ms{\Lambda}]$, and $\omega\in\Omega$, $\gamma_\Lambda(B,\omega)=\1{B}(\omega)$.
  \end{description}
\end{definition}

Properties (i) and (ii) are the usual definition of \emph{probability kernel} from $\Omega$ to $\Omega_{ \pc{\Lambda}}$.
Property (ii) says that the kernels carry no information on the future of $\Lambda$, as is the case for the transition probabilities of PCAs or other stochastic processes. 
Property (iv) expresses the fact that the past and the outer time of $\Lambda$ are frozen;
randomness is present only inside $\Lambda$.
In Gibbsian theory, kernels with the latter property are called \emph{proper kernels}.
Property (iii) gives the kernels its oriented character: the randomness within $\Lambda$ depends only on its past.

The kernels are interpreted as conditional ---or transition--- probabilities on $\Lambda$ given the past (of the POCs).
Therefore, we will indistinctly denote them $\gamma_\Lambda(\cdot,\cdot)$ or $\gamma_\Lambda(\cdot|\cdot)$.  

\begin{definition}
  \label{def=POS}
  A \dfn{partially oriented specification} (\dfn{POS}) $\gamma$ on $(\Omega, \F)$ is a family of proper oriented kernels $\{\gamma_\Lambda\}_{\Lambda\in\Sb}$ such that
  \begin{description}
  \item[(v)]
    For all $\Lambda, \Delta\in\Sb$ such that $\Lambda\subset\Delta$, 
    \begin{equation}
      \label{eq:r1}
      \iint h(\xi)\ \gamma_\Lambda(d\xi,\sigma)\ \gamma_\Delta(d\sigma,\omega) \;=\; \int h(\sigma)\ \gamma_\Delta(d\sigma,\omega)
    \end{equation}
    for each $\F[\pc{\Lambda}]$-measurable bounded function $h$ and each configuration $\omega\in\Omega$, 
  \end{description}
\end{definition}

This property is usually termed \dfn{consistency} and summarily written in the form 
\[
\gamma_\Delta\gamma_\Lambda\;=\;\gamma_\Delta\qquad \mbox{on } \F[\pc{\Lambda}]\;,
\]
where the left-hand side is interpreted in the sense of composition (or convolution) of kernels.
In words, \reff{eq:r1} means that integrating a function on $\Lambda$ and then integrating the result on $\Delta$ is exactly the same as integrating the function directly on $\Delta$. 
In probabilistic terms, this means that $\gamma_\Lambda$ is the (regular) conditional probability of $\gamma_\Delta$ given $ \F[\pc{\Lambda}]$ and, hence, that the family $\{\gamma_\Delta :\Delta\in\Sb\}$ is a consistent family of regular conditional probabilities.  
The central issue is to find measures that realize, or ``explain'', these conditional probabilities.

\begin{definition}
  \label{dfn:POC}
  A probability measure $\mu$ on $(\Omega,\F)$ is said to be \dfn{consistent with a POS} $\gamma$ if for each $\Lambda\in\Sb$, 
  \begin{equation}
    \label{eq:r2}
    \iint h(\xi)\ \gamma_\Lambda(d\xi,\sigma)\ \mu(d\sigma) \;=\; \int h(\sigma)\ \mu(d\sigma)
  \end{equation}
  for each $\F[\pc{\Lambda}]$-measurable bounded function $h$.  
  Such a measure $\mu$ is called a $\gamma-$ \dfn{partially oriented chain} or a $\gamma-$ \dfn{POC}.
  The set of $\gamma-$POCs will be denoted by $\G(\gamma)$.
\end{definition}

Conditions \reff{eq:r2}, which  can be more briefly written as
\begin{equation}
  \mu\gamma_\Lambda\;=\;\mu\quad \qquad \mbox{on } \F[\pc{\Lambda}]\;,
\end{equation}
correspond to the DLR equations of statistical mechanics~\cite{Georgii}.  
They are the $\Delta\to S$ limit (``infinite-volume'' or ``thermodynamic'' limit) of the consistency condition \reff{eq:r1} and are equivalent to demanding that each $\gamma_\Lambda$ be the conditional expectation of $\mu$ (restricted to $\F[\pc{\Lambda}]$) given the past of $\Lambda$.
Thus, we have a typically statistical mechanical situation: Data ---the model--- comes in the form of a family of conditional probabilities and the problem is to find measures that realize them.
The objective of the theory is to make a catalog of consistent measures and their properties.
Borrowing standard statistical mechanical nomenclature, we will sometimes refer to consistent measures as \dfn{phases}.
In particular we may refer to \dfn{phase coexistence} if $\bigl|\G(\gamma)\bigr|>1$.

Let us now define important particular classes of POMs which are the analogous of well studied classes of processes and fields.

\begin{definition}
  \label{def:nearest past}
  Let $x\in S$ and $\Lambda\in\Sb$.
  \begin{itemize}
  \item[(i)]
    The \dfn{nearest past} of $x$ and $\Lambda$ are
    \begin{equation}
      \label{eqn:frontiere passée}
      \db x \; := \;\max(x_-) \qquad,\qquad
      \db \Lambda \; := \;\Bigl(\bigcup_{x\in\Lambda}\db x\Bigr)\prive\Lambda
    \end{equation}
  \item[(ii)]
    More generally, for $k\in\mathbb{N}$, the \dfn{$k$-past} of $x$ and $\Lambda$ are the sets $\db^kx$ and $\db^k\Lambda$ iteratively defined as $\db^k\Lambda \bydef \db(\db^{k-1}\Lambda)\cup\db^{k-1}\Lambda$.
  \end{itemize}
\end{definition}

Note that sites can be in different $k$-pasts. This implies that, in general, $\db^kx$ is not a time box (except for $\db x$).
See Figure~\ref{fig:k-past}.
\begin{figure}
  \begin{center}
    \input{Images/k-past.pstex_t}
    \caption{An example of geometry that shows that $\db^2x$ can be a bad box.
      Arrows indicates the partial order: they go from small sites to big sites.}
    \label{fig:k-past}
  \end{center}
\end{figure}

\begin{definition}
  A POS $\gamma$ is \dfn{local} if there exists $k\in\mathbb{N}$ such that for all $\Lambda\in\Sb$ and all $A\in\F[\Lambda]$, $\gamma_\Lambda(A|\cdot)$ is $\F[\db^k \Lambda]$-measurable. 
  In the literature, the term \dfn{partially ordered Markov model} (\dfn{POMM}) has been reserved for the case $k=1$, but the actual value of $k$ plays little role in the theory. 
  A probability measure on $(\Omega,\F)$ consistent with a POS of each of these types is called, respectively, a \dfn{partially ordered local chain} and \dfn{partially ordered Markov chain}.
\end{definition}

POMMs, or local POS, are the analogous of Markov chains for partially ordered ``time''.
Their natural generalization are  kernels depending on the whole past, but in a manner asymptotically insensitive to the farther past.
These are formalized in the definition that follows.

\begin{definition}
  \begin{itemize}
  \item[(i)]
    A measurable function $f$ is \dfn{quasilocal} if it is the uniform limit of local functions, or, equivalently, if for all $\varepsilon>0$ there exists a finite $\Lambda\subset S$ such that
    \[
    \sup_{\omega,\sigma:\omega_\Lambda=\sigma_\Lambda}\bigl|f(\omega) - f(\sigma)\bigr|\;<\; \varepsilon\;.
    \]
  \item[(ii)]
    A POS $\gamma$ is \dfn{quasilocal} if $\gamma_\Lambda(A,\cdot)$ is quasilocal for all local event $A$ and all time box $\Lambda$ such that $A\in\F[\pc{\Lambda}]$.
  \end{itemize}    
  A \dfn{partially ordered quasilocal chain} is a probability measure consistent with a quasilocal POS.
\end{definition}

In statistical mechanics, quasilocal specifications play a central role because they correspond to the Gibbs measures introduced in physics through interactions and Boltzmann weights.
We do not explore here the particular features of their partially ordered counterparts, except for the existence issue.
Indeed, when the color space $E$ is compact (for instance, finite) a simple compactness argument (Theorem \ref{th:existence} below) shows that every quasilocal POS has at least one consistent measure (obtained as the weak limit of some sequence $\gamma_{\Lambda_n}(\,\cdot\mid \omega^n)$ with $\Lambda_n\to S$).

%
%
\section{Results} 

In this section we summarize our results.
Proofs are presented in the sections that follow.
%
\subsection{Properties of kernels}

We begin with some elementary properties that follow directly from Definition \ref{def:POK}.

\begin{proposition}
  Let $\Lambda\in\Sb$ and $\gamma_\Lambda$ be a proper oriented kernel.
  \begin{itemize}
  \item[(i)] \label{prop:dans boite} 
    For all $A\in\F[\Lambda]$ and $B\in\F[\ms{\Lambda}]$,
    \[
    \gamma_\Lambda(A\cap B|\omega) = \1{B}(\omega)\gamma_\Lambda(A|\omega)
    \]
  \item[(ii)] \label{prop:lambda-}
    If $f$ is an $\F[\Upsilon]$-measurable  function for $\Upsilon\subset\pc{\Lambda}$, then $\gamma_\Lambda(f)\bydef\gamma_\Lambda(f,\cdot)$ is $\F[\Lambda_-\cup(\Upsilon\prive\Lambda)]$-measurable.
  \item[(iii)]   \label{prop:egalite}
    If $\bar{\gamma}_\Lambda$ is a proper oriented kernel on $\Lambda$ such that $ \gamma_\Lambda(f) = \bar{\gamma}_\Lambda(f)$ for all $\F[\Lambda]$-measurable functions $f$, then $\gamma_\Lambda = \bar{\gamma}_\Lambda$.
  \end{itemize}
\end{proposition}

\begin{remarques}\ \par
  \begin{itemize}
  \item[(a)]
    Part (i) shows that
    \begin{equation}\label{eq:rp5}
      \gamma_\Lambda(d\sigma | \omega) \;=\; \bar{\gamma}_\Lambda(d\sigma_\Lambda | \omega)\ \1{\omega_{\ms{\Lambda}}}(\sigma_{\ms{\Lambda}})
    \end{equation}
    where $\bar{\gamma}_\Lambda$ is a kernel on $\Lambda$.
    This explicitly shows that the past is indeed frozen.
    In the sequel, we shall use this property without distinguishing $\bar{\gamma}$ from $\gamma$.
  \item[(b)]    
    In particular, part(ii) implies that if $f$ is $\F[\Lambda\cup\Lambda_-]$-measurable then $\gamma_\Lambda(f)$ is $\F[\Lambda_-]$-measurable.
    In other words, no dependency of $\Lambda^*$ is added by applying $\gamma_\Lambda$ to $f$.
    What happens in $\Lambda^*$ does not influence what happens in $\Lambda$.
  \end{itemize}
\end{remarques}

For the next two results we suppose a countable color space $E$.
The first result shows that a POS is characterized by ---can be reconstructed from--- the single-site kernels.
The second result shows that, in fact, any family of single-site proper oriented kernels can be used to build a POS.

\begin{theoreme}[Reconstruction Theorem]
  \label{thm:reconstruction}
  Assume $E$ countable and consider a POS $\gamma$ and $\Delta\in\Sb$.
  Then, there exists a sequence $x_1,\cdots,x_n$ of the points of $\Delta$ such that
  \[
  \gamma_\Delta = \gamma_{x_1} \cdots \gamma_{x_n}
  \]
\end{theoreme}

The following theorem justifies the usual practice of defining POCs ---in particular PCA--- only through single-site kernels.

\begin{theoreme}[Construction Theorem]
  \label{thm:construction} Assume $E$ countable.
  For each family $(\gamma_x)_{x\in S}$ of single-site proper oriented kernels there exists a unique POS $\gamma=(\gamma_\Delta)_{\Delta\in\Sb}$ such that $\gamma_{\{x\}}=\gamma_x$ for all $x\in S$.
  Furthermore,
  \begin{equation}
    \label{eqn:G(gamma)}
    \G(\gamma)=\Big\{ \mu : \mu\gamma_x = \mu,\quad \text{for all }x\in S\Big\}\;.
  \end{equation}
\end{theoreme}

There is a conceptual difference between Theorem \ref{thm:reconstruction} and Theorem \ref{thm:construction}.
In the first one, we start with a POS and we reconstruct the kernel in $\Lambda$ with single-site kernels.
In the second one, we start with a family of single-site kernels and we construct a POS compatible with it.

The unconstrained freedom to define single-site kernels leading to POS puts the latter on an equal footing with discrete-time processes.
In contrast, single-site kernels for random fields need to satisfy some further compatibility conditions in order to give rise to full specifications~\cite{dacnah01,dacnah04,fermai03a,fernandez-maillard-2,dacnah08}.

%
\subsection{Properties of chains}

The theorems of this subsection show why, as for processes and statistical mechanical fields, interest focuses on chains that are extremal points of the convex set $\G$.  
Indeed, the following theorems show that these measures satisfy the following properties:
\begin{itemize}
\item[(a)]
  They are determined by the ``initial'' set-up on the (infinitely far away) past.
\item[(b)]
  They enjoy a very general mixing property: colors at far away sites behave almost independently. 
\item[(c)] 
  They behave deterministically on ``global'' observables.
\item[(d)]
  They can be ``locally seen'' in the sense that they can be approximated by finite-region kernels.
\end{itemize}
These are precisely the properties expected for physical ``macroscopic'' systems. 

\begin{figure}[htbp]
  \begin{center}
    \scalebox{0.5}{\input{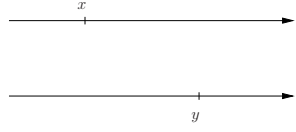}}
  \end{center}
  \caption{For two independent chains on $\ZZ$, the indicated sites $x$ and $y$ have $x_-\cap y_-=\emptyset$}
  \label{fig:Finfty}
\end{figure}

Our theorems refer to the $\sigma$-algebra
\begin{equation}
  \label{eqn:F-infty}
  \F[-\infty]:=\bigcap_{\Lambda\in\Sb}\F[\ms{\Lambda}]\;.
\end{equation}
Its elements can be roughly interpreted as events that do \emph{not} depend on any \emph{finite} family of sites.
This interpretation, however, has to be taken with a grain of salt.
Indeed, on the one hand the sites refer to exteriors of time boxes only and, on the other hand, the full complement of the future is involved.
In fact, the definition of $\F[-\infty]$ as $\bigcap_{\Lambda\in\Sb}\F[\Lambda_-]$ is unsuitable because it may happen that there exist $x,y\in S$ such that $x_-\cap y_-=\emptyset$ (see Figure \ref{fig:Finfty} for an example).
In this case, $\bigcap_{\Lambda\in\Sb}\F[\Lambda_-]=\{\Omega,\emptyset\}$.
Lemma \ref{lem:cap ms} below shows that Definition \reff{eqn:F-infty} never leads to such a trivial $\sigma$-algebra.  

Our results are summarized in three theorems.

\begin{theoreme}
  \label{thm:measure general prop}
  Let $\gamma$ be a POS on $(\Omega,\F)$.
  The following properties hold:
  \begin{description}
  \item[(a)]
    $\G(\gamma)$ is a convex set.
  \item[(b)]
    A measure $\mu$ is extremal in $\G(\gamma)$ if and only if it is trivial on $\F[-\infty]$.
  \item[(c)]
    Let $\mu\in\G(\gamma)$ and $\nu$ be a measure on $\F$ such that $\nu\ll\mu$.
    Then $\nu\in\G(\gamma)$ if and only if there exists a nonnegative $\F[-\infty]$-measurable function $h$ such that $\nu=h\mu$.
  \item[(d)]
    Each $\mu\in\G(\gamma)$ is uniquely determined [within $\G(\gamma)$] by its restriction to $\F[-\infty]$.
  \item[(e)]
    Two distinct extremal elements of $\G(\gamma)$ are mutually singular on $\F[-\infty]$.
  \end{description}
\end{theoreme}
\medskip

\begin{theoreme}
  \label{thm:extremal caracterisation}
  For each probability measure $\mu$ on $(\Omega, \F)$, the following statements are equivalent:
  \begin{description}
  \item[(a)]
    $\mu$ is trivial on $\F[-\infty]$.
  \item[(b)]
    For all cylinder sets $A\in\F$,
    \begin{equation}
      \label{eqn:ergodique avec cylindres}
      \lim_{\Lambda\uparrow S} \sup_{B\in\F[\ms{\Lambda}]} \big| \mu(A\cap B)-\mu(A)\mu(B) \big| = 0
    \end{equation}
  \item[(c)]
    For all $A\in\F$,
    \begin{equation}
      \label{eqn:ergodique}
      \lim_{\Lambda\uparrow S} \sup_{B\in\F[\ms{\Lambda}]} \big| \mu(A\cap B)-\mu(A)\mu(B) \big| = 0
    \end{equation}
  \end{description}
\end{theoreme}
\medskip

\begin{theoreme}
  \label{thm:extremal as limit}
  Let $\gamma$ be a POS, $\mu$ an extremal point of $\G(\gamma)$ and $(\Lambda_n)_{n\in\NN}$ a sequence of time boxes such that $\Lambda_n\uparrow S$.
  Then
  \begin{enumerate}
  \item[(i)]
    $\gamma_{\Lambda_n}(h)\rightarrow\mu(h)\ \mu$-a.s. for each bounded local function $h$ on $\Omega$.
  \item[(ii)]
    If $\Omega$ is a compact metric space, then for $\mu$-almost all $\omega\in\Omega$
    \begin{equation}\label{eq:rp1}
      \gamma_{\Lambda_n}(h\mid\omega) \xrightarrow[n\to\infty]{} \mu(h)
    \end{equation}
    for all continuous local functions $h$ on $\Omega$.
  \end{enumerate}
\end{theoreme}

Notice that in part (i) the set of full measure where convergence takes place can, in general, be different for different $h$.
In contrast, in (ii) there is a full-measure set where the convergence holds simultaneously for all local continuous $h$.
This last convergence can be interpreted as the possibility to understand an extremal measure by observing kernels in big but finite boxes with a typical past condition.
In particular, this feature holds for models with finite color space $E$.

We conclude with an existence theorem for quasilocal POS.

\begin{theoreme}
  \label{th:existence}
  \noindent
  Let $\gamma$ be a quasilocal POS.
  \begin{itemize}
  \item[(i)]
    Let $(\nu_n)$ be a sequence of probability measures on $\Omega$ and $(\Lambda_n)$ a sequence of time boxes such that there exists a probability measure $\mu$ with
    \begin{equation}
      \label{eq:rp2}
      \lim_n \int \nu_n(d\omega)\,\gamma_{\Lambda_n}(h\mid\omega) \;=\; \mu(h)
    \end{equation}
    for all continuous local functions $h$ on $\Omega$.
    Then $\mu\in\G(\gamma)$.
  \end{itemize}
  If, in addition, $\Omega$ is separable and compact,
  \begin{itemize}
  \item[(ii)]  
    $\G(\gamma)\neq\emptyset$.
  \item[(iii)]
    If there exist a local function $h$ and two configurations $\omega_1,\omega_2\in\Omega$ such that
    \[
    \lim_{n\to\infty}\gamma_{\Lambda_n}(h|\omega_1) \neq \lim_{n\to\infty}\gamma_{\Lambda_n}(h|\omega_2)
    \]
    then, $\bigl|\G(\gamma)\bigr|\ge 2$ \ \ \ ($\gamma$ exhibits ``phase coexistence'').
  \end{itemize}
\end{theoreme}

%
\subsection{Inequalities related to color ordering}

We group under this heading a number of results derived from the presence of a total order on the color space $E$.
We fix a total order for $E$ and consider the induced partial order on $\Omega$:
\[
\omega\leq\eta \qquad \ssi \qquad \forall x\in S,\ \omega_x\leq\eta_x\;.
\]
[We use the same symbol for the orders on $S$, on $\Omega$ and on $E$;
the context determines which one applies.]
An order leads to the associated notions of increasing and decreasing functions, as well as a corresponding notion at the level of measures.

\begin{definition}
  Let $\mu$ and $\nu$ be two probability measures on the same (partially) ordered space $X$.
  We say that $\mu$ is \dfn{stochastically dominated} by $\nu$ and we denote $\mu\dom\nu$ if $\mu(f)\leq\nu(f)$ for all non-decreasing bounded measurable functions $f$.
\end{definition}

The main results in this section are summarized in the following theorem, which is a transcription to the POS setting  of the FKG (Fortuin-Kasteleyn-Ginibre) inequalities, a well known tool in statistical mechanics.

\begin{theoreme}[FKG]
  \label{thm:FKG}
  Let $\gamma$ be a POS such that for all $y\in S$, $a\in E$ and $\omega,\eta\in\Omega$ with $\omega\leq\eta$,
  \begin{equation}\label{eq:rfkg}
    \gamma_y\Big(\sigma_y\geq a\Bigm| \omega\Big) \;\leq\; \gamma_y\Big(\sigma_y\geq a\Bigm| \eta  \Big)\;.
  \end{equation}
  Then it satisfies the following FKG inequalities:
  \begin{itemize}
  \item[(i)]
    For all $\Lambda\in\Sb$, 
    \begin{equation}\label{eq:fkg.0}
      \gamma_\Lambda(\cdot\mid\omega) \;\dom\; \gamma_\Lambda(\cdot\mid\eta) \quad \mbox{on }\F[\pc{\Lambda}]\;.
    \end{equation}
  \item[(ii)]
    For all local increasing functions $f,g$, for all $\Delta\in\Sb$ such that $\Supp(f)$ and $\Supp(g)$ are in $\pc{\Delta}$ and for all $\omega\in\Omega$,
    \begin{equation}
      \label{eqn:FKG-local}
      \gamma_\Delta\Big(fg\Bigm|\omega\Big) \geq \gamma_\Delta\Big(f\Bigm|\omega\Big) \gamma_\Delta\Big(g\Bigm|\omega\Big)\;.
    \end{equation}
  \item[(iii)] 
    For all extremal measures $\mu$ in $\G(\gamma)$ and all increasing functions $f$, $g$,
    \begin{equation}
      \label{eqn:FKG-extremal}
      \mu(fg) \geq \mu(f)\mu(g)
    \end{equation}
  \end{itemize}
\end{theoreme}

These inequalities provide a very powerful tool for the study of POCs.
The following theorem summarizes the most common form of exploiting them.

\begin{theoreme}\label{th:color-ordering}
  Consider a quasilocal POM with a color-ordering such that:
  \begin{itemize}
  \item[(a)]
    $\max E=\{u\}$ and $\min E=\{d\}$ for some colors $u,d\in E$.
  \item[(b)]
    The model satisfies hypothesis \reff{eq:rfkg} of the previous theorem.  
  \end{itemize}
  Let us denote $\oplus$, resp.\ $\ominus$, the ``all up'', resp.\ ``all-down'', configurations ($\oplus_x=u$, $\ominus_x=d$ for all $x\in S$).
  Then,
  \begin{itemize}
  \item[(i)]
    For every time box $\Lambda$ and every configuration $\omega$, 
    \begin{equation}
      \gamma_\Lambda(\cdot\mid\ominus) \;\dom\; \gamma_\Lambda(\cdot\mid\omega) \;\dom\; \gamma_\Lambda(\cdot\mid\oplus)\quad \mbox{on }\F[\pc{\Lambda}]\;.
    \end{equation}
  \item[(ii)]
    For all time boxes $\Lambda,\Delta$ such that $\Lambda\subset\Delta$ and $\Lambda_+\cap\Delta=\emptyset$, 
    \begin{equation}
      \gamma_\Delta(f\mid\oplus) \;\leq\; \gamma_\Lambda(f\mid\oplus)
    \end{equation}
    for all increasing $\F[\Lambda]$-measurable functions $f$.
  \item[(iii)]
    The weak limits
    \begin{equation}\label{eq:rp20}
      \mu^\oplus := \lim_{\Lambda\uparrow S}\gamma_\Lambda(\cdot\mid\oplus) \quad,\quad
      \mu^\ominus := \lim_{\Lambda\uparrow S}\gamma_\Lambda(\cdot\mid\ominus)    
    \end{equation}
    exist and belong to $\G(\gamma)$.
  \item[(iv)]
    $\mu^\oplus$ and $\mu^\ominus$ are the only extremal $\gamma$-POC.
  \item[(v)]
    $\bigl|\G(\gamma)\bigr| = 1$ if, and only if, $\mu^\oplus=\mu^\ominus$. 
  \end{itemize}
\end{theoreme}

In particular, the theorem applies to our benchmark examples.

\begin{proposition}\label{pro:ising-stra}
  The POMM-Ising and Stavskaya models introduced in Section \ref{ss:benchmark} above satisfy hypothesis (a) and (b) of Theorem \ref{th:color-ordering} with $u=1$ and $d=-1$ (Ising) or $d=0$ (Stavskaya).
  Therefore conclusions (i)--(v) of the theorem hold for these models.
\end{proposition}

In fact, with a little more work one can prove that for both the Ising and Stavskaya models
\begin{equation}
  \mu^\oplus = \mu^\ominus \quad \ssi \quad \mu^\oplus(\sigma_{(0,0)}) = \mu^\ominus(\sigma_{(0,0)})\;.
\end{equation}
The proof is an adaptation of \cite{Lebowitz-Martin-lof}.
As a consequence, in both cases, the uniqueness of the consistent POC is equivalent to the condition $\lim_n \gamma\bigl(\sigma_{(0,0)}\bigm| \ominus) = \lim_n \gamma\bigl(\sigma_{(0,0)}\bigm| \oplus)$.
In particular, for the Stavskaya model $\mu^\ominus$ is just the Dirac measure concentrated in the ``all zero'' configuration.
Then, there is phase coexistence if, and only if, there exists a consistent measure $\nu$ such that $\nu(\sigma_{(0,0)})>0$.

%
\subsection{Uniqueness criteria}

Since $\G$ is convex, its cardinal can take only three values: 0, 1 or infinity.  
The theorems of this subsection determine conditions for this cardinal to be at most 1.
They apply to progressively more restricted set-ups:
The bounded-uniformity criterion refers to general POMs (though it is useful only in a few), Dobrushin's requires a countable color space and disagreement percolation demands, in addition, Markovianness. 

\subsubsection{Bounded uniformity}

Our first theorem is the transcription of a theorem used in statistical mechanics to prove that one-dimensional finite-range models do not exhibit phase coexistence (see, e.g.~\cite{Georgii}, Section 8.3).
\begin{theoreme}
  \label{thm:useless}
  Let $\gamma$ be a POS for which there exists a constant $c>0$
  such that for all cylinders $A$ there exists a time box $\Lambda$
  such that $A\in\F[\pc{\Lambda}]$ and
  \begin{equation}
    \label{eqn:useless critere}
    \forall\omega, \xi\in\Omega\quad \gamma_\Lambda(A,\omega)
    \geq c\ \gamma_\Lambda(A,\xi)
  \end{equation}
  Then $\big|\G(\gamma)\big|\leq1$.
\end{theoreme}
This is not a very useful criterion.  
For local specifications it can be applied only when the number of nearest-past sites of time boxes remains bounded as the box grows.

\subsubsection{Dobrushin criterion}

We present the version useful for a countable color space $E$.
Generalizations are possible for metrizable $E$, but we focus on the simplest version for the sake of clarity.
The criterion results from a beautiful inductive argument to ``clean'' oscillations of conditioned averages.
Its formalization requires a few introductory definitions.

For $\xi, \eta\in\Omega$ and $x\in S$, let us write $\xi\pequ{x}\eta$ if $\xi_y=\eta_y$ for all $y\in S\prive\{x\}$.

\begin{definition}
  Let $f:\Omega\mapsto\RR$ be a measurable function.
  \begin{itemize}
  \item[(i)]
    The {\bf oscillation} of $f$ with respect to the site $x$ is
    \begin{equation}
      \label{eqn:delta_s}
      \delta_x(f) := \sup_{\xi\pequ{x}\eta}\big| f(\xi)-f(\eta) \big|
    \end{equation}
  \item[(ii)]
    The {\bf total oscillation} of $f$ is
    \begin{equation}
      \label{eqn:Delta}
      \Delta(f) := \sum_{x\in S}\delta_x(f)
    \end{equation}
  \end{itemize}
\end{definition}
Note that every bounded local function $f$ has bounded total oscillation and, furthermore,
\begin{equation}
  \label{eqn:Delta continu}
  \sup(f)-\inf(f) \;\leq\; \Delta(f)\; <\; \infty\;.
\end{equation}

\begin{definition}
  \label{def:crm}
  A {\bf dust-rate matrix} $(\alpha_{y,x})_{x\leq y}$ is a matrix of nonnegative real numbers such that for all $y\in S$, $\alpha_{y,y}=0$  and 
  \begin{equation}
    \label{eqn:dust rate matrix}
    \delta_x(\gamma_yf)\;\leq\;\delta_y(f)\ \alpha_{y,x}
  \end{equation}
  for all $x\in y_-$ and all $\F[\{y\}]$-measurable functions $f$.
\end{definition}
By Proposition \ref{prop:lambda-} there is no need to define $\alpha_{y,x}$ for $x\in \ps{y}$.
[Alternatively, we can set $\alpha_{y,x}:=0$ if $x\in \ps{y}$.]

The name ``dust-rate matrix'' comes from an interpretation due to Aizenman:
Imagine that $S$ is a tiling of an infinite room and associate oscillations of functions to dust. 
The application of the kernel $\gamma_y$ to $f$ produces a new function $\gamma_yf$ that has no ``dust'' at $y$ (it no longer depends on the color at $y$).
Thus, $\gamma_y$ can be thought as a broom that perfectly cleans the site $y$.
However, the oscillations of $\gamma_yf$ at sites in $y_-$ will be different form the original oscillations of $f$.
This fact can be attributed to dust thrown out by the broom during the cleaning of $y$.
The coefficient $\alpha_{y,x}$ represents the maximal rate of dust that can be sent from $y$ to $x$. 

With this interpretation, uniqueness can be associated to the existence of a ``cleaning procedure'' that successively removes dust from all sites, producing conditioned averages with less and less oscillations.
Weak limits of these averages become, therefore, insensitive to external conditions and all lead to the same unique consistent measure.
For such a  program to have a chance to succeed, each application of a broom must do some actual cleaning, that is, the dust that flies away must be less than that that was removed.
In more quantitative terms, the total dust rate must be less than one.
Dobrushin criterion proves that such a condition indeed implies uniqueness.

\begin{theoreme}
  \label{thm:Dobrushin}
  Let $\gamma$ be a quasi-local POS on a countable color space $E$.
  If there exists a dust-rate matrix $\alpha$ such that
  \begin{equation}
    \label{eqn:Dobrushin critere}
    \Gamma\;\bydef\;\sup_{y\in S}\sum_{x\leq y}\alpha_{y,x}\;<\;1
  \end{equation}
  then there exists at most one $\gamma$-POC.
\end{theoreme}

Of course, the efficiency of this criterion crucially depends on a good estimation of the dust-rate matrix $\alpha$.
The following proposition provides a reasonable estimate.

\begin{proposition}
  \label{prop:crm2s}
  Consider a POS $\gamma$ with countable color space.
  Then, the numbers
  \begin{equation}\label{eq:rp30}
    \alpha_{y,x} \;=\; \sup_{\xi\pequ{x}\eta}\;\frac12 \sum_{a\in E}\Bigl| \gamma_y(a\mid\xi) - \gamma_y(a\mid\eta) \Bigr|
  \end{equation}
  define a dust-rate matrix.
  Furthermore, if $\left|E\right|=2$ these are the smallest possible entries for a dust-rate matrix.  
\end{proposition}

In fact, if $\left|E\right|=2$, say $E=\{d,u\}$, the expression becomes
\begin{equation}\label{eq:rp30.2}
  \alpha_{y,x} \;=\; \sup_{\xi\pequ{x}\eta}\Bigl| \gamma_y(u\mid\xi) - \gamma_y(u\mid\eta) \Bigr|
  \;=\; \sup_{\xi\pequ{x}\eta}\Bigl| \gamma_y(d\mid\xi) - \gamma_y(d\mid\eta) \Bigr|\;.
\end{equation}
Educated readers may have recognized that the right-hand side of \reff{eq:rp30} involves the variational distance between the measures $\gamma_y(\cdot\mid\xi)$ and $\gamma_y(\cdot\mid\eta)$ projected on $\Omega_y$.
This is the root of a number of extensions and generalization of the criterion that we prefer not to develop here.

%
\subsubsection{Oriented disagreement percolation}

This criterion requires Markovianness, thus we will be dealing with POMMs.
Also, the color space $E$ is assumed to be countable, though generalizations are possible.
The criterion is based on the distribution of the sites where two coupled realizations differ.
Uniqueness ensues if a coupling can be found such that these disagreement sites do not percolate.
Let us first present the oriented percolation set-up relevant for our models.

\begin{definition}
  Consider a partially ordered set $(S,\le)$ and a family of parameters $\pg = (p_x)_{x\in S}$ with each $p_x\in[0,1]$.
  \begin{itemize}
  \item[(i)]
    Let $\psi_{\textbf{p}}$ denote the independent Bernoulli distribution on $S$ with parameters $\pg$. 
    Let $X\in \{0,1\}^S$ denote a random variable with this distribution.
    A site $x$ is \dfn{open} if $X_x=1$, event that happens with probability $p_x$.
    If $p_x=q$ for all $x\in S$ the distribution is denoted $\psi_q$.  
  \item[(ii)]
    For $y<x\in S$ let
    \begin{equation}
      (x\join[>] y)\;=\; \Bigl\{X\in\{0,1\}^S:\ \exists (x_k)_{1\leq k\leq n} \mbox{ with }x_1=x,\ x_n=y,\ x_k\in\db x_{k+1},\ X_{x_k}=1\Bigr\}\;.
    \end{equation}
    This is the event ``there exists an oriented (towards the past) path from $x$ to $y$''.
    (We recall that $\db x$ is the nearest past of the site $x$, see Definition \ref{def:nearest past}.)
  \item[(iii)]
    A site $x\in S$ belongs to an \dfn{infinite oriented 1-cluster}, denoted by $(x\join[>]-\infty)$ if there exists an infinite decreasing sequence $(y_n)_{n\in\NN}$ such that $(x\join[>]y_n)$ for all $n\in\NN$.
  \item[(iv)]
    The distribution $\psi_{\textbf{p}}$ \dfn{percolates} if $\psi_{\textbf{p}}(x\join[>] -\infty)>0$.
  \item[(v)]
    The \dfn{critical oriented percolation parameter} of $S$ is the value 
    \begin{equation}
      p_c^+ \;:=\; \inf\Bigl\{q:  \psi_q(x\join[>] -\infty)>0\Bigr\}\;.
    \end{equation}
  \end{itemize}
\end{definition}

We remark that by the POS-Holley Theorem presented below (Theorem~\ref{thm:POS-Holley}), $\psi_q\dom\psi_{q'}$ whenever $q<q'\in[0,1]$.
Hence the critical percolation parameter can be equivalently defined as $p_c^+ \;=\; \sup\bigl\{q: \psi_q(x\join[>] -\infty)=0\bigr\}$.

The disagreement criterion is based on Bernoulli percolation with parameters derived from the POMM in the following way. 

\begin{definition}
  The \dfn{maximal percolation parameters} of a POMM $\gamma$ is the family $\pg^\gamma = (p_x^\gamma)_{x\in S}$ defined by
  \begin{equation}
    \label{eqn:p_s}
    p_x^\gamma\;=\;\sup_{\eta,\xi\in\Omega} \frac12 \sum_{a\in E}\Bigl| \gamma_y(a\mid\eta) - \gamma_y(a\mid\xi) \Bigr|\;.
  \end{equation}
\end{definition}
[The right-hand side involves a variational distance, as in Proposition \ref{prop:crm2s}.]
Note that, in particular, if $E=\{u,d\}$ has only two colors, 
\begin{equation}
  \label{eqn:p_s-2}
  p_x^\gamma \;=\; \sup_{\eta,\xi\in\Omega} \big| \gamma_x(u\mid\eta)-\gamma_x(u\mid\xi) \big| \;=\; \sup_{\eta,\xi\in\Omega} \big| \gamma_x(d\mid\eta)-\gamma_x(d\mid\xi) \big|\;.
\end{equation}

We can finally state the criterion.

\begin{theoreme}
  \label{thm:dp criterion}
  A POMM $\gamma$ on a countable color space $E$ has a unique consistent chain if the distribution $\psi_{\textbf{p}^\gamma}$ does not percolate.
\end{theoreme}
In practice, this criterion is applied in the following form. 
\begin{corollaire}
  \label{coro:dp criterion}
  A POMM $\gamma$ on a countable color space $E$ has a unique consistent chain if
  \begin{equation}
    \label{eqn:coupling critere}
    \sup_{x\in S}p_x^\gamma < p_c^+\;.
  \end{equation}
\end{corollaire}

\begin{figure}[htbp]
  \begin{center}
    \includegraphics[width=3cm]{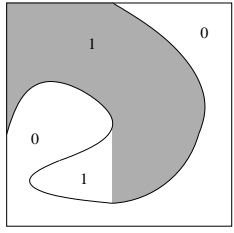}
  \end{center}
  \caption{oriented 1-cluster (in grey) versus non-oriented 1-cluster in $\ZZ^2$ with our drawing conventions.}
  \label{fig:1-cluster}
\end{figure}

There are two aspects that determine in practice the efficiency of this criterion.
First, we need a good estimation of the parameters $p_x^\gamma$.
For civilized models, like our benchmark examples, this is not a complicated task, and for more involved models one can resort to calculator- or computer-assisted evaluations.
A more delicate second aspect is the determination of the critical parameter $p_c^+$ for the oriented set $(S,\le)$, a step that leads to models not usually studied in percolation theory.
However, oriented percolation is more restrictive than unoriented percolation because clusters have to obey more constraints in the former (Figure \ref{fig:1-cluster} shows an example).
Therefore $p_c^+$ is not smaller than its unoriented counterpart $p_c$ and one can always, at least as a first approximation, apply \reff{eqn:coupling critere} using the better known value $p_c$ instead of $p_c^+$. 

%
%
\section{Application: Uniqueness in the benchmark examples}
\sectionmark{Application}

As an illustration, let us apply Dobrushin and oriented disagreement percolation criteria to our benchmark examples.

\subsection{The POMM-Ising model}
\label{subsec:uniq pomm-ising}

\paragraph{Dobrushin criterion.}
Let $\alpha_{y,x}$ be the dust-rate matrix \reff{eq:rp30.2}.
Since each $\I_y$ depends only on the sites $Ny$ and $Wy$, $\alpha_{y,z}=0$ for all $z\in S\prive\{Ny,Wy\}$.
By symmetry $\alpha_{y,Wy}=\alpha_{y,Ny}$.
Furthermore, for $\eta\in\Omega$, denote $\eta^{Ny,1}$ the configuration $\eta$ where we impose $\eta_{Ny} = 1$. We define similarly $\eta^{Ny,-1}$.
\[\begin{split}
  \alpha_{y,Ny} & = \sup_{\eta\in\Omega} \Big| \I_y(1,\eta^{Ny,1}) - \I_y(1,\eta^{Ny,-1}) \Big| \\
  & = \frac{\sinh(2\beta)}{\cosh(2\beta)+\cosh(2\beta(|h|-1))} \\
\end{split}\]
The Dobrushin criterion gives uniqueness of the $\I$-POC for
\begin{equation}
  \label{eqn:dobrushin Pomm-Ising}
  \sum_{x: x\le y }\alpha_{y,x}\; =\; 2\,\alpha_{y,Ny} \;=\; \frac{2\sinh(2\beta)}{\cosh(2\beta)+\cosh(2\beta(|h|-1))}\;<\;1
\end{equation}
In particular, this shows that there is uniqueness for all $\beta>0$ if the external field $h$ is equal to zero. 
That is, the voter model in $\ZZ^2$ never exhibits phase coexistence.
In fact, in a companion paper \cite{pomm-geom} (see also \cite{deveaux-thesis}) uniqueness is shown to hold, through an expansion-based approach, also for $h\neq 0$.  

\paragraph{Oriented disagreement percolation.}
The maximal oriented percolation parameters \reff{eqn:p_s-2} are
\[\begin{split}
  p_x^\I & = \sup_{\eta,\xi\in\Omega} \big| \I_x(1,\eta) - \I_x(1,\xi) \big| \\[8pt]
  & = \frac{1}{2}\Bigl[\tanh\bigl(\beta(|h|+2)\bigr) - \tanh\bigl(\beta(|h|-2)\bigr)\Bigr]\;. \\
\end{split}\]
Since $p_c^+(\ZZ^2)>\frac{1}{2}$, the disagreement-percolation criterion proves uniqueness at least if 
\begin{equation}
  \label{eqn:DP POMM-Ising}
  \tanh\bigl(\beta(|h|+2)\bigr) - \tanh\bigl(\beta(|h|-2)\bigr)\; <\; 1\;.
\end{equation}
A not negligible improvement is obtained by using the more accurate value $p_c^+\approx 0.64450$ obtained by Monte Carlo methods~\cite{belitsky}.
It corresponds to replacing 1 by 1.289 in \reff{eqn:DP POMM-Ising}.
\bigskip

\begin{figure}[htbp]
  \begin{center}
    \scalebox{0.5}{\input{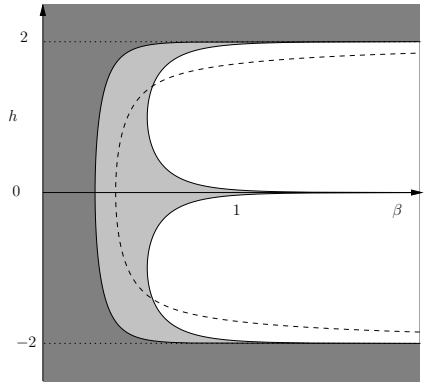}}
  \end{center}
  \caption{The dark grey region corresponds to the parameters for which both Dobrushin and oriented disagreement percolation criteria (with 1/2 instead of $p_c^+$) can be applied.
    In the light grey zone, only Dobrushin criterion is valid.
    The dashed line corresponds to the frontier of the region obtained with a numerically accurate value of $p_c^+$.
    In the white zone, none of the criteria can be applied.}   
  \label{fig:Pomm-Ising}
\end{figure}

Figure \ref{fig:Pomm-Ising} summarizes the different results.
While Dobrushin criterion is more efficient than disagreement percolation with 1/2 as lower bound for $p_c^+$, the latter gives stronger results if the numerical value is used for $p_c^+$.
There is also a large region of the $(\beta,h)$-plane where none of our criteria gives information.

\subsection{The Stavskaya's model}
\label{subsec:uniq stav}

As for the preceding model, the only non-zero entries of the dust-rate matrix \reff{eq:rp30.2} are
\begin{equation}
  \label{eqn:Dobrushin percolation}
  \alpha_{y,Wy}\;=\; \alpha_{y,Ny} \;=\; \sup_{\eta_{Wy}\in\{0,1\}} \big| p\1{\eta_{Wy}>0} - p\1{1+\eta_{Wy}>0} \big| \;=\; p\;.
\end{equation}
Hence, Dobrushin criterion proves uniqueness if $2p < 1$.

The maximal oriented percolation parameters \reff{eqn:p_s-2} are
\begin{equation}
  \label{eqn:percolation or. perco.}
  p_x \;= \sup_{\eta_{Nx},\eta_{Wx},\xi_{Nx},\xi_{Wx}\in\{0,1\}} \big| p\1{\eta_{Nx}+\eta_{Wx}>0} - p\1{\xi_{Nx}+\xi_{Wx}>0} \big| \;=\; p
\end{equation}
Disagreement percolation leads, therefore, to uniqueness for $p<p_c^+$, which is a an improvement over Dobrushin's.
In fact, this condition can be seen to be optimal: For $p>p_c^+$ there are at least two $\S$-POC \cite{pomm-geom,deveaux-thesis}.

This example can easily be generalized to any lattice with finite (oriented) neighborhood.
If such a lattice has $n$ past neighbors per site then $p_c^+ \leq 1/n$.
This bound, however, is independent of the geometry of the lattice and, for instance, gives the same result for $\ZZ^2$ with its natural partial order and for the infinite binary tree.  

We observe that the combination of the Dobrushin and disagreement percolation results shows the well known lower bound $p_c^+ \geq 1/2$.

%
%
\section{POMM versus PCA and Gibbs field}
\label{sec:POMM vs PCA}

%
\subsection{POMM versus PCA}

Before starting with the formal proofs let us briefly discuss the differences between POMMs and PCAs.
We shall justify the assertion:
\begin{equation}
  \mbox{Every PCA is a POMM but the converse is false}
\end{equation}

Let us first agree on the definition of probability cellular automata.
The ingredients of the standard definition~(see \cite{toom-impa}) are as follows:
\begin{itemize}
\item
  A countable set $U$ of sites.
\item
  For each $i\in U$, a finite set $V_i\subset U$ containing $i$, called \dfn{the neighborhood} of $i$.
\item
  A measurable space $(E,\mathcal{E})$ (spin values, occupation numbers, \ldots) defining what is usually called the space of \dfn*{(spatial) configurations} $(\widetilde{\Omega}=E^U,\mathcal{E}^U)$.
  We warn the reader that what we have called configurations would correspond to space-time configurations in PCAs. 
\item
  A family of single-site probability kernels $\theta_i$ from $\widetilde{\Omega}$ to $(E,\mathcal{E})$ interpreted as transition probabilities.
  For each $y\in\widetilde\Omega$ and $A_i\in \mathcal{E}$, the value of $\theta_i(A_i\mid y)$ represents the probability of falling into $A_i$ having started from $y$.
  Furthermore, $\theta_i(A_i\mid \cdot) $ is $\mathcal{E}^{V_i}$-measurable.
  As customary, let us stress this fact by writing $\theta_i(A_i\mid y_{V_i}) $.
\item
  A transition probability kernel on $\widetilde\Omega$ defined by
  \begin{equation}
    \label{eqn:PCA}
    P(dx\mid y) = \prod_{i\in U} \theta_i(dx_i\mid y_{V_i})\;.
  \end{equation}
  This corresponds to a ``parallel updating'' of configurations:
  Conditionally on $y$, what happens at each site is independent of what happens at all other sites.
\end{itemize}
Iterations of the stochastic transformation \reff{eqn:PCA} define a discrete-time stochastic process;
the orders of iteration defining a ``time'' axis identified with $\NN$.
The iterations are started on some initial distribution (often concentrated on a single configuration) and interest focuses on the invariant measures of the dynamics.
These measures are, in principle defined by the consistency condition $\mu P = \mu$ (c.f.\ Definition~\ref{dfn:POC}), but in reasonable cases should also be attainable as the limit of infinitely many iterations of the dynamics.
The process can also be defined on $\ZZ$, shifting the time $0$ to time $-n$ and letting $n\to\infty$.
In the case of phase coexistence, however, such a limit can depend on the initial distribution.
If this is taken as one of the invariant measures the resulting process on $\ZZ$ is invariant under time shifts.

\begin{figure}[htbp]
  \begin{center}
    \includegraphics[width=5cm]{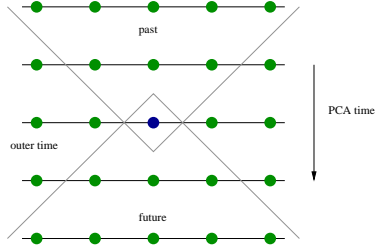}
  \end{center}
  \caption{PCA seen as a POMM: here $U=\ZZ$ and $V_i=\{i-1,\ i,\ i+1\}$.
    The central site depends on the three sites above it.}
  \label{fig:PCA as POMM}
\end{figure}

The canonical way to write a PCA as a POMM is along  the following lines:
\begin{itemize}
\item[(i)]
  The site space is $S:=U\times\ZZ$ with the partial order where the past of a point $(i,t)$ is formed by all points that have contributed to the transitions leading to it.
  Formally, 
  \[
  (i,t) \leq (j,s) \quad \ssi \quad
  \begin{cases}
    (i,t) = (j,s) \\
    \qquad \text{or} \\
    \exists (k_n)_{0 \leq n \leq s-t}\in U^{s-t+1}\ k_0=j,\ k_{s-t}=i,\ k_n\in V_{k_{n+1}} \\
  \end{cases}
  \]
\item[(ii)]
  The POMM is defined ---via the construction theorem and identity \reff{eq:rp5}--- by the single-site kernels
  \[
  \gamma_{(i,t)}(d\sigma_i | \eta) := \theta_i(d\sigma_{(i,t)} | \eta_{V_i, t-1})
  \]
  for every $i\in U$ and $t\in\ZZ$.
\end{itemize}
Figure \ref{fig:PCA as POMM} gives an idea of the construction.
It is straightforward to check that $\gamma$ is a POMM with slices of the form $(V,t):=\{(i,t)\in S;\ i\in V\}$ for $V\subset U$, $t\in\ZZ$.
Therefore,
\[
\begin{split}
  P(x_V | y_{t-1}) & = \prod_{i\in V}\theta_i(x_i\mid y_{V_i,t-1}) \\
  & = \prod_{i\in V}\gamma_{(i,t)}(x_i \mid y) \\
  & = \gamma_{(V,t)}(x_V \mid y) \\
\end{split}
\]
[see Corollary \ref{corol:produit} below].
\medskip

We see that a PCA is a particular type of POMM with an order given by a product structure (``flat slices'').
The following is an example of a POMM that can not be written as PCA. 
\begin{exemple}
  \label{exem:pomm-pas-pca}
  Let $S:=\ZZ\times\{0\} \cup \{0\}\times\NN$.
  We define a partial order on $S$ by:
  \[
  (x,i) \leq (y,j) \quad \ssi \quad
  \begin{cases}
    i=j=0,\ x \leq y \\
    \qquad \text{or} \\
    i=y=0,\ x \leq j \\
    \qquad \text{or} \\
    x=j=0,\ i \leq y \\
  \end{cases}
  \]
  The geometry of $S$ is shown in Figure \ref{fig:pomm non pca}.
  It can not be written as a product of space because of the shortcuts created by $\{0\}\times\NN$.
  \begin{figure}[htbp]
    \begin{center}
      \includegraphics[width=5cm]{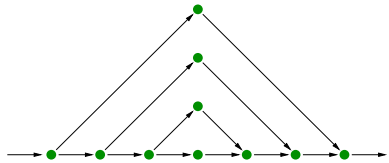}
    \end{center}
    \caption{An example of geometry that shows that a POMM can not be a PCA.
      Arrows indicates the partial order:
      they go from small sites to big sites.}
    \label{fig:pomm non pca}
  \end{figure}
\end{exemple}

Incidently, either the bounded uniformity or the disagreement percolation criteria prove that every non-null POMM based on the geometry of Example \ref{exem:pomm-pas-pca} has $\left|\G(\gamma)\right|=1$.
For instance, a Bernoulli field $\Psi_q$ can not percolate in this graph unless $q=1$ for the same reason that this is impossible in $\ZZ$.
Thus, the oriented disagreement percolation criterion leads to the conclusion that if $\gamma_x(e|\omega) >0$ for all $x\in S$, $\omega\in\Omega$ and $e\in E$, then there is only one $\gamma$-POC.  

\subsection{POMM versus Gibbs specifications}

In this section we link POMMs and Gibbs fields.
The latter are consistent with (unoriented) specifications defined by conditioning on the \emph{whole} exterior being frozen.
Such specification is Gibbsian if kernels are non-null (definition follows) and become asymptotically independent of far away sites.
In particular, non-null Markovian specifications ----namely those whose kernels depend only on neighboring exterior sites--- are Gibbsian.
A Gibbs measure or field is a measure consistent with a Gibbsian specification (see~\cite{Georgii} for precise definitions).        
The following arguments show that, in the presence of non-nullness, Markovian partially ordered chains are Markovian Gibbs fields but the converse is not always true.

We start with two definitions.

\begin{definition}
  Let $\gamma$ be a POMM, $x\in S$ and $\Lambda\in\Sb$.
  The \dfn{nearest future} of $x$ and $\Lambda$ are
  \begin{equation}
    \ol{\partial}x := \min(x_+) \quad , \qquad
    \ol{\partial}\Lambda := \left( \bigcup_{x\in\Lambda}\ol{\partial}x \right) \prive\Lambda
  \end{equation}
\end{definition}

\begin{definition}
  A POMM $\gamma$ is \dfn{non-null} if $\forall x\in S$, $\forall \eta\in\Omega$, $\forall e\in E$, $\gamma_x(e|\eta)>0$.
\end{definition}

Note that, by the Reconstruction theorem, $\forall \Lambda\in\Sb$, $\forall \eta\in\Omega$, $\forall \sigma_\Lambda\in\Omega_\Lambda$, $\gamma_\Lambda(\sigma_\Lambda|\eta)>0$.

Consider now a POMM $\gamma$, $\Upsilon$ be a finite part of $S$ and $\Lambda\in\Sb$ such that $\big( \db \Upsilon \cup \ol{\partial} \Upsilon \cup \Upsilon \big) \subset \Lambda$.
For $\sigma_\Lambda\in\Omega_\Lambda$ and $\eta\in\Omega$,
\begin{align}
  \label{eq:gibbs}
  P_\Upsilon(\sigma_\Upsilon \mid \sigma_{\Lambda\prive\Upsilon}, \eta) & := \frac{\gamma_\Lambda(\sigma_\Upsilon \sigma_{\Lambda\prive\Upsilon} \mid \eta)} {\gamma_\Lambda(\sigma_{\Lambda\prive\Upsilon} \mid \eta)} \nonumber \\[8pt]
  & = \frac{\prod_{x\in\Upsilon} \gamma_x(\sigma_x\mid\sigma_{\Lambda\prive\{x\}}\eta) \prod_{x\in\ol{\partial}\Upsilon}\gamma_x(\sigma_x\mid\sigma_\Lambda\eta)} {\sum_{\tilde{\sigma}_\Upsilon\in\Omega_\Upsilon}\prod_{x\in\Upsilon} \gamma_x(\tilde{\sigma}_x\mid\sigma_{\Lambda\prive\{x\}}\eta) \prod_{x\in\ol{\partial}\Upsilon}\gamma_x(\sigma_x\mid\tilde{\sigma}_\Lambda\eta)} \nonumber \\[8pt]
  & = \frac{1}{Z} \prod_{x\in\Upsilon}\gamma_x(\sigma_x\mid\sigma_{\Lambda\prive\{x\}}\eta) \prod_{x\in\ol{\partial}\Upsilon}\gamma_x(\sigma_x\mid\sigma_\Lambda\eta)
\end{align}
Here $Z$ is a normalizing coefficient independent of $\sigma_\Upsilon$.
The markovianness of $\gamma$ implies that the LHS is independent of $\eta$.
Therefore \reff{eq:gibbs} defines a Markovian (unoriented) specification.
It is easy to check that every measure consistent with $\gamma$ is also consistent with the specification defined by \reff{eq:gibbs}.
This shows that every partially ordered Markovian chain is a Gibbs field.

For example, the Gibbs field associated with the POMM Ising model is defined on singletons by
\[
P(\sigma_x|\eta) = \frac{1}{Z} \frac{\e^{\beta\sigma_x(\eta_{Nx}+\eta_{Wx}+\eta_{Sx}+\eta_{Ex})}} {\Big( \e^{\beta\eta_{Sx}(\sigma_x+\eta_{SWx})} + \e^{\beta\eta_{Sx}(-\sigma_x+\eta_{SWx})} \Big)  \Big( \e^{\beta\eta_{Ex}(\sigma_x+\eta_{NEx})} + \e^{\beta\eta_{Ex}(-\sigma_x+\eta_{NEx})} \Big)}
\]
$Sx$, $Ex$, $SWx$ and $NEx$ correspond respectively to the southern, eastern, south-western and north-eastern site of $x$.
Note that the interactions between $x$ and $Nx$, $Wx$, $Ex$ and $Sx$ are ferromagnetic whereas the interaction between $x$ and $SWx$ and $NEx$ are anti-ferromagnetic.
In particular, this shows that the associated Gibbs field of the POMM Ising is not the Ising model.

\begin{figure}[htbp]
  \begin{center}
    \input{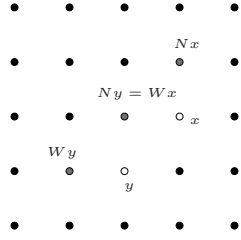}
  \end{center}
  \caption{Configuration that shows that the Ising model is not a POMM.}
  \label{fig:PommIsing-nonGibbs}
\end{figure}

In fact, the Ising model provides an example of a Markovian Gibbs field that can not be consistent with a POMM.
Indeed, if it were, the sites would be independent upon fixing the configuration on sites $Wy$, $Ny$ and $Nx$ (see Figure \ref{fig:PommIsing-nonGibbs}).
This is not the case of the Ising model.

%
%
\section{Proofs of the properties of kernels}

%
\subsection{Proof of Proposition \protect\ref{prop:dans boite}}
\subsectionmark{Proposition\protect\ref{prop:dans boite}}

\begin{preuve}[Proof of (i)]
  Fix $B\in\F[\ms{\Lambda}]$.
  For $\omega\not\in B$, $\gamma_\Lambda(A\cap B|\omega)=0$ and for $\omega\in B$, $\gamma_\Lambda(A\cap B|\omega) \leq \gamma_\Lambda(A|\omega)$.
  Then for all $A\in\F[\Lambda]$, we can write $\gamma_\Lambda(A\cap B | \omega) \leq \gamma_\Lambda(A | \omega)\ \1{B}(\omega)$.
  Now, by the ``proper'' character of $\gamma_\Lambda$ [part (iv) of Definition \ref{def:POK}],
  \[
  0 = \Big( \gamma_\Lambda(A\cap B|\omega) - \gamma_\Lambda(A | \omega) \1{B}(\omega) \Big) + \Big( \gamma_\Lambda(A^c\cap B|\omega) - \gamma_\Lambda(A^c | \omega) \1{B}(\omega) \Big)\;.
  \]
  This proves part (i) because both terms are non-positive.
\end{preuve}

\begin{preuve}[Proof of (ii)]
  Let $\Xi=\Lambda_-\cup(\Upsilon\prive\Lambda)$ and $\xi,\eta\in\Omega$ such that $\xi_\Xi=\eta_\Xi$.
  Let us define $f_\eta:\Omega\mapsto\RR$ by $f_\eta(\omega):=f(\omega_\Lambda\eta_{\Lambda^c})$.
  Since $f_\eta$ is $\F[\Lambda]$-measurable, $\gamma_\Lambda(f_\eta)$ is $\F[\Lambda_-]$-measurable by part (iii) of Definition \ref{def:POK}.
  This implies that $\gamma_\Lambda(f_\eta,\xi) = \gamma_\Lambda(f_\eta,\eta)$.
  Moreover, by part (i), $\gamma_\Lambda(f) = \gamma_\Lambda(f_\eta)$.
  So we have
  \[ \begin{split}
    \gamma_\Lambda(f,\xi)-\gamma_\Lambda(f,\eta) & = \gamma_\Lambda(f_\xi,\xi) - \gamma_\Lambda(f_\eta,\eta) \\
    & = \gamma_\Lambda(f_\xi,\xi) - \gamma_\Lambda(f_\eta,\xi) \\
    & = \gamma_\Lambda(f_\xi-f_\eta,\xi) \\
    & = 0\;. \\
  \end{split} \]
  The last line is due to the fact that $f_\eta=f_\xi$ because $f$ is $\F[\Upsilon]$-measurable.
\end{preuve}

\begin{preuve}[Proof of (iii)]
  Let $g$ be an $\F[\pc{\Lambda}]$-measurable function and $\eta$ be a configuration.
  We have to prove that $\gamma_\Lambda(g)(\eta) = \bar{\gamma}_\Lambda(g)(\eta)$.
  As in the proof of part (ii), let $g_\eta$ denote the function defined by $g_\eta(\xi) := g(\xi_\Lambda\eta_{\Lambda^c})$.
  The function $g_\eta$ is $\F[\Lambda]$-measurable, thus
  \[
  \gamma_\Lambda (g,\eta) = \gamma_\Lambda(g_\eta,\eta) = \bar{\gamma}_\Lambda(g_\eta,\eta) = \bar{\gamma}_\Lambda (g,\eta)\;.
  \]
\end{preuve}

%
\subsectionmark{Reconstruction theorem}
\subsection{Proof of the reconstruction theorem}
\label{sec:rebuild}
\subsectionmark{Reconstruction theorem}

%
\subsubsection{Slicing}

The reconstruction scheme is based on a procedure that we call \emph{slicing}.
To define it we need a number of properties of kernels on time boxes.
We prove these for general time boxes but later will be used mostly for single-site boxes.

\begin{proposition}
  \label{prop:rond passé}
  Let $\Lambda, \Delta\in\Sb$ such that $\Delta\cap\Lambda=\emptyset$, $\Delta\cap\Lambda_+=\emptyset$ and $\Lambda_-\cap\Delta_+=\emptyset$.
  Let $\gamma_\Lambda$ and $\gamma_\Delta$ be proper oriented kernels on $\Lambda$ and $\Delta$.
  Denote $\Gamma:=\Lambda\cup\Delta$.
  Then,
  \begin{itemize}
  \item[(i)]
    $\Gamma\in\Sb$, and $\Gamma_+=\Lambda_+\cup(\Delta_+\prive\Lambda)$, $\Gamma_-=\Delta_-\cup(\Lambda_-\prive\Delta)$,
  \item[(ii)]
    $\gamma_\Gamma:=\gamma_\Delta\gamma_\Lambda$ is well defined and is a proper oriented kernel on $\Gamma$.
  \end{itemize}
\end{proposition}

\begin{figure}[htbp]
  \begin{center}
    \scalebox{0.5}{\input{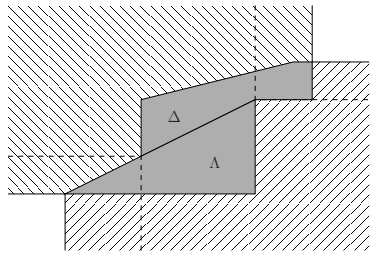}}
  \end{center}
  \caption{Typical use of Proposition \ref{prop:rond passé}}
  \label{fig:reconstr-ordonne}
\end{figure}

\begin{preuve}[Proof of (i)]
  We show first that $\Lambda\cap\Delta_-=\emptyset$:
  Indeed, if there exist $x\in\Delta_-\cap\Lambda$, there would exist $y\in\Delta$ such that $y>x$.
  But this would imply a contradiction because $y\in\Lambda_+\cap\Delta=\emptyset$.
  As a consequence,
  \begin{equation}
    \label{eq:r5}
    \Lambda_+\;=\;\Lambda_+\prive\Delta \quad,\qquad
    \Delta_-\;=\;\Delta_-\prive\Lambda\;.
  \end{equation}
  Let us denote $U_\Delta:=\bigcup_{t\in\Delta}t_-$ and similarly for $U_\Lambda$ and $U_\Gamma$.
  We see that $\Delta_-=U_\Delta\prive\Delta$ and likewise for $\Lambda_-$ and $\Gamma_-$.
  Hence
  \[\begin{split}
    \Delta_-\cup(\Lambda_-\prive\Delta) & = (\Delta_-\prive\Lambda) \cup (\Lambda_-\prive\Delta) \\
    & = \big( U_\Delta\prive\Gamma \big) \cup \big( U_\Lambda\prive\Gamma \big) \\
    & = U_\Gamma\prive\Gamma \\
    & = \Gamma_- \\
  \end{split}\]
  The proof that $\Lambda_+\cup\big(\Delta_+\prive\Lambda\big)=\Gamma_+$ is analogous.
  
  To prove that $\Gamma\in\Sb$ we have to show that $\Gamma_+\cap\Gamma_-=\emptyset$.
  Our previous relations show that
  \[\begin{split}
    \Gamma_+\cap\Gamma_- & = \Big( \Lambda_+\cup\big(\Delta_+\prive\Lambda\big) \Big) \cap \Big( \Delta_-\cup(\Lambda_-\prive\Delta) \Big) \\
    & = (\Lambda_+ \cap \Delta_-) \cup \Big(\Lambda_+ \cap (\Lambda_-\prive\Delta)\Big) \cup \Big((\Delta_+\prive\Lambda) \cap \Delta_-\Big) \cup \Big((\Delta_+\prive\Lambda) \cap (\Lambda_-\prive\Delta)\Big) \\
    & = \Lambda_+ \cap \Delta_- \;.\\
  \end{split}\]
  If $\Gamma_+\cap\Gamma_-\neq\emptyset$, there would exist $x\in\Lambda_+\cap\Delta_-$.
  But this leads to a contradiction because it implies the existence of $y\in\Delta$ such that $x<y$ and thus $y\in\Delta\cap\Lambda_+$.
  This intersection is, however, empty by the middle identity in \ref{eq:r5}.
\end{preuve}

\begin{preuve}[Proof of (ii)]
  The proof involves three verifications.
  \smallskip
  
  \noindent \emph{$\gamma_\Gamma$ is well defined}.
  Since $\Gamma_+ = \Lambda_+\cup(\Delta_+\prive\Lambda)$, we have $\pc{\Gamma}\subset\pc{\Lambda}$, so we can apply $\gamma_\Lambda$ on any $A\in\F[\pc{\Gamma}]$.
  By part (ii) of Proposition \ref{prop:lambda-}, $\gamma_\Lambda(A)$ is $\F[\Upsilon]$-measurable, where $\Upsilon=\Lambda_-\cup[\pc{\Gamma}\prive\Lambda]=\Lambda_-\cup\ms{\Gamma}\cup\Delta$.
  Moreover, $\Upsilon\cap\Delta_+ = (\Lambda_-\cap\Delta_+) \cup (\ms{\Gamma}\cap\Delta_+) = \emptyset$ (the first intersection is empty by hypothesis).
  Thus, it is possible to apply $\gamma_\Delta$ to $\gamma_\Lambda(A)$.
  The function $\gamma_\Gamma(A)$ is then well defined and is $\F[\Upsilon']$-measurable where $\Upsilon'=\Delta_-\cup[\Upsilon\prive\Delta]=\Gamma_-\cup\ms{\Gamma}=\ms{\Gamma}$.
  \smallskip
  
  \noindent \emph{$\gamma_\Gamma$ is oriented}.
  Indeed, by the previous result, for any $A\in\F[\Gamma]$, the function $\gamma_\Gamma(A)$ is $\F[\Upsilon_1]$-measurable where $\Upsilon_1=\Delta_-\cup[[\Lambda_-\cup(\Gamma\prive\Lambda)]\prive\Delta] =\Delta_-\cup[(\Lambda_-\cup\Delta)\prive\Delta] =\Delta_-\cup(\Lambda_-\prive\Delta) = \Gamma_-$.
  \smallskip

  \noindent \emph{$\gamma_\Gamma$ is proper}. 
  Let $B\in\F[\ms{\Gamma}]$.
  Since
  \[ \begin{split}
    \ms{\Gamma}\cap(\Lambda_+\cup\Lambda) & = \ms{\Gamma} \cap \Lambda_+ \\
    & \subset \ms{\Gamma} \cap (\Lambda_+\cup(\Delta_+\prive\Lambda)) \\
    & \subset \ms{\Gamma}\cap\Gamma_+ \\
    & = \emptyset\;, \\
  \end{split} \]
  we have $\ms{\Gamma}\subset\ms{\Lambda}$ so $\gamma_\Lambda(B) = \1{B}$ and $\gamma_\Lambda(B)$ is $\F[\ms{\Gamma}]$-measurable.
  Finally,
  \[ \begin{split}
    \ms{\Gamma}\cap(\Delta_+\cup\Delta) & = \ms{\Gamma}\cap\Delta_+ \\
    & = \ms{\Gamma}\cap(\Delta_+\prive\Lambda \\
    & \subset \ms{\Gamma} \cap (\Lambda_+\cup(\Delta_+\prive\Lambda)) \\
    & \subset \ms{\Gamma} \cap\ \Gamma_+ \\
    & = \emptyset\;.
  \end{split} \]
  Hence $\gamma_\Gamma(B)=\gamma_\Delta(B)=\1{B}$.
\end{preuve}

\begin{corollaire}
  \label{prop:commute}
  Let $\Delta, \Lambda\in\Sb$ such that $\Delta\subset\Lambda^*$ and denote $\Gamma:=\Delta\cup\Lambda$.
  Let $\gamma_\Delta$ and $\gamma_\Lambda$ be proper oriented kernels on $\Delta$ and $\Lambda$.
  Then,
  \begin{itemize}
  \item[(i)]
    $\Gamma\in\Sb$, and $\Gamma_+=\Delta_+\cup\Lambda_+$, $\Gamma_-=\Delta_-\cup\Lambda_-$,
  \item[(ii)]
    $\gamma_\Gamma:=\gamma_\Delta\gamma_\Lambda$ is well defined and is a proper oriented kernel on $\Gamma$,
  \item[(iii)]
    If $E$ is countable, $\gamma_\Gamma=\gamma_\Delta\gamma_\Lambda=\gamma_\Lambda\gamma_\Delta$.
  \end{itemize}
\end{corollaire}

\begin{figure}[htbp]
  \begin{center}
    \scalebox{0.6}{\input{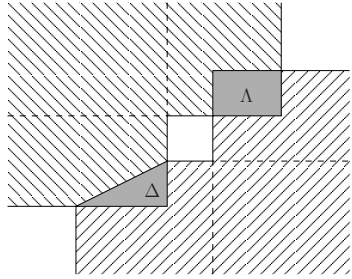}}
  \end{center}
  \caption{Typical use of Corollary \ref{prop:commute}}
  \label{fig:reconstr-indep}
\end{figure}

\begin{preuve}
  The previous proposition proves parts (i) and (ii).
  Note that we can exchange the role of $\Delta$ and $\Lambda$, hence $\bar{\gamma}_\Gamma:=\gamma_\Lambda\gamma_\Delta$ is a well-defined oriented kernel.
  By part (iii) of Proposition \ref{prop:egalite} it remains to prove that $\gamma_\Gamma(h) = \bar{\gamma}_\Gamma(h)$ for every $\F[\Gamma]$-measurable function $h$.
  This identity is true if $h=fg$ with $f$ is  $\F[\Delta]$-measurable and $g$ $\F[\Lambda]$-measurable, because
  \[
  \gamma_\Gamma(fg) = \gamma_\Delta(f)\gamma_\Lambda(g) = \bar{\gamma}_\Gamma(fg)\;.
  \]
  The equality for general $h$ follows from the decomposition
  \[
  h \;= \sum_{\substack{\sigma_\Delta\in E^\Delta \\ \sigma_\Lambda\in E^\Lambda}} h(\sigma_\Delta\sigma_\Lambda)\1{\sigma_\Delta}\1{\sigma_\Lambda}\;.
  \]
\end{preuve}

These two results give an easy way to construct proper oriented kernels on unrelated and ordered sets.
We shall apply them to families of sites called \emph{slices}.
\begin{definition}
  \label{def:slice}
  A finite set $\Lambda\subset S$ is a \dfn{slice} if all points of $\Lambda$ are pairwise unrelated.
\end{definition}

\begin{proposition}
  \label{prop:Sb}
  \begin{itemize}
  \item[(i)]
    A slice is a time box.
  \item[(ii)]
    for each finite subset $\Upsilon$ of $S$, $\max(\Upsilon)$ and $\min(\Upsilon)$ are slices.
  \item[(iii)]
    Each finite subset of $S$ is contained in the past of a time box.
  \end{itemize}
\end{proposition}
\begin{preuve}
  \noindent \emph{(i)}
  Let $\Delta$ be a slice and assume there exists $x\in\Delta_-\cap\Delta_+$.
  Then, there exist $y,z\in\Delta$ such that $y<x$ and $x<z$.
  So $y\leq z$, which is absurd by definition of $\Delta$.
  \smallskip\par
  \noindent \emph{(ii)}
  This is a simple consequence of the definitions of $\max$, $\min$ and slice.
  \smallskip\par
  \noindent \emph{(iii)}
  Let $\Upsilon$ be a finite subset of $S$.
  For each $x\in\max(\Upsilon)$ choose some $y>x$ and denote $\Delta$ the set of these $y$ (it can happen that $\Delta\cap\Upsilon\neq\emptyset$).
  The set $\max(\Delta)$ is the time-box we look for.
\end{preuve}

\begin{definition}
  \label{def:slice-decomp}
  Let $\Delta\in\Sb$ and define the following sequence of slices:
  \begin{equation}
    \label{eqn:strates}
    \begin{split}
      \Delta_1 & :=  \min(\Delta) \\
      \Delta_2 & :=  \min\left(\Delta\prive\Delta_1\right) \\
      & \vdots  \\
      \Delta_k & := \min\big(\Delta\prive(\Delta_1\cup\cdots\cup\Delta_{k-1})\big) \\
      & \vdots  \\
    \end{split}
  \end{equation}
  The \dfn{slicing of $\Delta$} is the sequence  $(\Delta_1,\cdots,\Delta_n)$ where $n$ is the greatest integer such that $\Delta_n\neq\emptyset$.
  The nonempty $\Delta_i$ are the \dfn{slices} of $\Delta$. 
\end{definition}

\begin{figure}[htbp]
  \begin{center}
    \includegraphics[height=3cm]{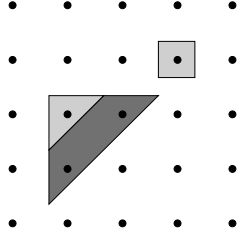}
  \end{center}
  \caption{$\Delta_1$ is in light grey, $\Delta_2$ in dark grey.
    $\Delta_2\neq\max(\Delta)$. }
  \label{fig:slice max}
\end{figure}

\begin{remarque}
  In general, $\Delta_n$ is not equal to $\max(\Delta)$ but it is a subset of it.
  Figure \ref{fig:slice max} shows an example of such a case.
\end{remarque}

%
\subsubsection{Proof of Theorem \protect\ref{thm:reconstruction}}

The proof has two parts:
First the box is split into slices, then each slice is split into sites.
\begin{preuve}
  Let $(\Delta_1,\cdots,\Delta_n)$ be the slicing of $\Delta$ and define
  \begin{equation}
    \bar{\gamma}_\Delta = \gamma_{\Delta_1}\gamma_{\Delta_2}\cdots\gamma_{\Delta_n}
  \end{equation}
  By Proposition \ref{prop:rond passé}, $\bar{\gamma}_\Delta$ define
  a proper oriented kernel on $\pc{\Delta}$.
  Indeed, the definition of slices implies that the sets
  $\Lambda_k = \cup_{i=k+1}^n\Delta_i$, $1\leq k \leq n-1$ satisfy
  $\Lambda_k\subset(\Delta_k)_+\subset \Lambda_k\cup\Delta_+$ so
  $(\Lambda_k)_+\cap\Delta_k=\emptyset$ and $(\Lambda_k)_-\cap(\Delta_k)_+=\emptyset$.
  Thus, the sets $\Lambda_k $ satisfy the hypotheses of Proposition \ref{prop:rond passé}.
  
  The fact that $\bar{\gamma}_\Delta=\gamma_\Delta$ is a consequence of
  the consistency property of the POS.
  Indeed, applying $n$ times property (v) in Definition \ref{def=POS}, we have that
  for all $\F[\pc{\Delta}]$-measurable functions~$f$
  \begin{equation}\label{eq:r6}
    \begin{split}
      \gamma_\Delta(f) & = \gamma_\Delta\left(\gamma_{\Delta_n}(f)\right) \\
      & = \cdots \\
      & = \gamma_\Delta\left(\gamma_{\Delta_1}\cdots\gamma_{\Delta_n}(f)\right) \\
      & = \gamma_\Delta\left(\bar{\gamma}_\Delta(f)\right) \\
      & = \gamma_\Delta(1)\bar{\gamma}_\Delta(f) \\
      & = \bar{\gamma}_\Delta(f) \;. \\
    \end{split}
  \end{equation}
  Now, if $\Lambda=\{y_1\cdots y_m\}$ be a slice, then by Corollary \ref{prop:commute}, $\gamma_{y_1}\cdots\gamma_{y_m}$ is a proper oriented kernel on $\Lambda$.
  Consistency implies
  \begin{equation}\label{eq:r7}
    \gamma_\Lambda = \gamma_\Lambda\big(\gamma_{y_1}\cdots\gamma_{y_m}\big) = \gamma_{y_1}\cdots\gamma_{y_m}\;.
  \end{equation}
  The combination of \reff{eq:r6} and \reff{eq:r7} proves the theorem.
\end{preuve}

We conclude emphasizing that by Corollary \ref{prop:commute}, the order of the single-site kernels within a slice is not important.

\begin{corollaire}
  \label{corol:produit}
  Let $\gamma$ be a POS.
  Then for all $\Delta\in\Sb$ and $\eta\in\Omega$,
  \[
  \gamma_\Delta(\sigma|\eta) = \prod_{x\in\Delta} \gamma_x(\sigma_x|\sigma_\Delta\eta_{\ms{\Delta}})\;.
  \]
\end{corollaire}

%
\subsectionmark{Construction theorem}
\subsection{Proof of the construction theorem}
\subsectionmark{Construction theorem}

\begin{preuve}[Construction of the POM]  
  Let $\Delta$ be in $\Sb$.
  If it is a slice, define
  \[
  \gamma_\Delta := \gamma_{y_1}\cdots\gamma_{y_m}
  \]
  where $\Delta=\{y_1,\cdots,y_m\}$.
  This is well defined according to Corollary \ref{prop:commute}.
  Otherwise, let $(\Delta_1,\cdots,\Delta_n)$ be the slicing of $\Delta$ and define
  \[
  \gamma_\Delta := \gamma_{\Delta_1}\cdots\gamma_{\Delta_n}
  \]
  By proposition \ref{prop:rond passé}, $\gamma_\Delta$ is a well defined proper oriented kernel.
  \smallskip\par
  
  \noindent{\sl Proof of consistency.}
  
  The rest of the proof relies on the following observation, valid for any measure $\mu$ on $\F$ and any $\Delta\in\Sb$:
  \begin{equation}
    \label{eqn:proof trans}
    \Big[ \forall x\in\Delta,\ \mu\gamma_x=\mu \Big] \quad \Longrightarrow \quad \mu\gamma_\Delta=\mu
  \end{equation}
  This is an immediate consequence of the fact that $\gamma_\Delta$ is obtained as the iteration of single-site kernels.  
  \smallskip
  
  This observation directly implies \reff{eqn:G(gamma)}.
  The proof of uniqueness of the POS $\gamma$ is only slightly less trivial.     
  Indeed, consider any other POS $(\bar{\gamma}_\Lambda)_{\Lambda\in\Sb}$ consistent with the family $(\gamma_x)_{x\in S}$.
  By \reff{eqn:proof trans}, $\bar{\gamma}_\Lambda$ must be consistent with $\gamma_\Lambda$ for each $\Lambda\in\Sb$.
  But then, if $f$ is $\F[\pc{\Lambda}]$-measurable
  \[
  \bar{\gamma}_\Lambda(f) = \bar{\gamma}_\Lambda\big(\gamma_\Lambda(f)\big) = \gamma_\Lambda(f)\bar{\gamma}_\Lambda(1) = \gamma_\Lambda(f)\;.
  \]
  \smallskip
  
  To conclude the theorem let us prove consistency of the kernels $\gamma_\Delta$.
  Let $\Lambda\in\Sb$ such that $\Lambda\subset\Delta$.
  To prove that $\gamma_\Delta\gamma_\Lambda = \gamma_\Delta$ it is enough, by \reff{eqn:proof trans}, to prove that  $\gamma_\Delta\gamma_y = \gamma_\Delta$ for each $y\in\Lambda$.  
  Pick such a $y\in\Lambda$ and consider any $\F[\pc{y}]$-measurable function $h$.
  Temporarily denote $f:=\gamma_y(h)$.
  Let $p$ be such that $y\in\Delta_p$.
  Since $\ms{y}\subset \ms{x}$ for each $x\in\Delta_{p+1}\cup\dots\cup\Delta_n$, 
  \begin{equation}
    \label{eq:r10}
    \gamma_\Delta\gamma_y(h) = \gamma_\Delta(f) = \gamma_{\Delta_1}\cdots\gamma_{\Delta_n}(f) = \gamma_{\Delta_1}\cdots\gamma_{\Delta_p}(f)\;.
  \end{equation}
  Assume now that $\Delta_p:=\{y_1,\cdots,y_m\}$.
  Since the inter-slice order is irrelevant, we can suppose that $y_m=y$.
  In this case,
  \begin{equation}
    \label{eq:r11}
    \gamma_{\Delta_p}(f)=\gamma_{y_1}\cdots\gamma_{y_m}\gamma_y(h) = \gamma_{\Delta_p}(h)\;.
  \end{equation}
  From \reff{eq:r10} and \reff{eq:r11} we conclude that
  \begin{equation}
    \label{eq:r12}
    \gamma_\Delta\gamma_y(h) = \gamma_{\Delta_1}\cdots\gamma_{\Delta_p}(h) = \gamma_\Delta(h)\;.
  \end{equation}
\end{preuve}

%
%
\section{Proofs of the properties of POCs}

\subsection{Proof of Theorem \protect\ref{thm:measure general prop}}

We begin with a result that in particular implies that $\F[-\infty]$ is not trivial.

\begin{lemme}
  \label{lem:cap ms}
  For all $x,y\in S$, $\ms{x}\cap \ms{y}\neq\emptyset$.
\end{lemme}
\begin{preuve}
  The proof is by contradiction.
  Assume there exist $x$ and $y$ such that $\ms{x}\cap \ms{y}=\emptyset$.
  Then $\ms{x}\subset S\prive \ms{y}=y_+\cup\{y\}$ and any $z\in x_-\subset y_+\cup\{y\}$, satisfies  that $z\leq x$ and $y\leq z$.
  This implies that $y\leq x$ and hence $y_-\subset x_-$.
  This contradicts the original assumption that $\ms{x}\cap \ms{y}=\emptyset$.
\end{preuve}
\medskip

The proof of the theorem is based on the following two lemmas taken from \cite[pages 115-117]{Georgii}.
\begin{lemme}
  Let $(\Omega, \B)$ be a measurable space, $\pi$ a probability kernel from $\B$ to $\B$ and $\mu$ a measure on $\B$ such that $\mu\pi=\mu$.
  Denote
  \[
  \A_\pi^\B(\mu):=\Bigl\{A\in\B, \pi(A,\cdot)=\1{A}(\cdot)\ \mu\text{-a.s.}\Bigr\}
  \]
  Then, $\A_\pi^\B(\mu)$ is a $\sigma$-algebra and for every $\B$-measurable nonnegative function h,
  \[
  \Big( (h\mu)\pi=h\mu \Big) \ssi \Big( h\text{ is }\A_\pi^\B(\mu)\text{-measurable} \Big).
  \]
\end{lemme}
\smallskip

\begin{lemme}
  Let $(\Omega,\B)$ be a measurable space and $\Pi$ a non-empty set of kernels such that, for all $\pi\in\Pi$, $\pi$ is a probability kernel from $\B_\pi$ to $\B$, where $\B_\pi$ is a sub-$\sigma$-algebra of $\B$.
  Denote
  \[
  \G(\Pi):=\Bigl\{\mu\in\P(\Omega, \B): \mu\pi=\mu\,\ \forall\pi\in\Pi\Bigr\}
  \]
  the convex set of $\Pi$-invariant probability measures and for $\mu\in\G(\Pi)$, $\A_\Pi(\mu):=\bigcap_{\pi\in\Pi}\A_\pi^{\B_\pi}(\mu)$.
  Then
  \[
  \Big( \mu\text{ is extremal in }\G(\Pi) \Big) \ssi \Big( \mu\text{ is trivial on }\A_\Pi(\mu) \Big)
  \]
\end{lemme}

\begin{preuve}[Proof of Theorem \ref{thm:measure general prop}]
  \noindent \emph{(a)}
  \ \ Its proof is immediate.
  \medskip\par
  \noindent \emph{(b)}
  \ \ Denote $\F[-\infty]^\mu$ the $\mu$-completion of $\F[-\infty]$.
  We only have to prove that $\F[-\infty]^\mu=\A_\gamma(\mu)$, because $\mu$ is trivial on $\F[-\infty]$ if and only if $\mu$ is trivial on $\F[-\infty]^\mu$.
  
  Let $A\in\A_\gamma(\mu)$.
  For each time box $\Lambda$, $A\in\A_{\gamma_\Lambda}^{\F[\pc{\Lambda}]}$ so $\gamma_\Lambda(A,\cdot)=\1{A}(\cdot)\ \mu$-a.s..
  Then, $\1{A}$ is $\mu$-a.s. $\F[\ms{\Lambda}]$-measurable, that is, $A\in\F[\ms{\Lambda}]$ $\mu$-a.s..
  This implies that $A\in\F[-\infty]^\mu$.
  
  Conversely, let $A\in\F[-\infty]^\mu$.
  Thus, there exists a set $B\in\F[-\infty]$ such that $A=B\ \mu$-almost surely.
  Let $\Lambda$ be a time box.
  Since $B\in\F[\ms{\Lambda}]$ for all $\Lambda\in\Sb$, $\gamma_\Lambda(B,\cdot)=\1{B}(\cdot)$ and $B\in\A_{\gamma_\Lambda}^{\F[\pc{\Lambda}]}$.
  Thus, $A\in\A_{\gamma_\Lambda}^{\F[\pc{\Lambda}]}(\mu)$ for each time box $\Lambda$, which proves that $A\in\A_\gamma(\mu)$.
  
  \medskip\par
  \noindent \emph{(c)}
  \ \ $\nu\ll\mu$ implies that there exists an $\F$-measurable non-negative function $f$ such that~$\nu=f\mu$.
  \begin{eqnarray*}
    \nu\in\G(\gamma) & \ssi & \forall\Lambda\in\Sb,\ \nu\gamma_\Lambda=\nu \\
    & \ssi & \forall\Lambda\in\Sb,\ (f\mu)\gamma_\Lambda=f\mu \\
    & \ssi & \forall\Lambda\in\Sb,\ f\text{ is } \A_{\gamma_\Lambda}^{\F[\pc{\Lambda}]}(\mu)\text{-measurable} \\
    & \ssi & f\text{ is } \A_\gamma(\mu)\text{-measurable} \\
    & \ssi & f\text{ is } \F[-\infty]^\mu\text{-measurable} \\
    & \ssi & \exists h\ \F[-\infty]\text{-measurable such that }h=f\ \mu\text{-a.s.}\\
    & \ssi & \exists h\ \F[-\infty]\text{-measurable such that }\nu=h\mu \\
  \end{eqnarray*}
  
  \medskip\par
  \noindent \emph{(d)}
  \ \ Let $\mu$, $\nu\in\G(\gamma)$ such that their restrictions to $\F[-\infty]$ coincide and define $\overline{\mu}:=\frac{\mu+\nu}{2}\in\G(\gamma)$.
  Since $\mu\ll\overline{\mu}$, there exists a $\F[-\infty]$-measurable function $f$ such that $\mu=f\overline{\mu}$.
  But $\mu=\overline{\mu}$ on $\F[-\infty]$, thus $f=1$ $\overline{\mu}$-a.s. and as a consequence $\mu=\overline{\mu}$.
  Analogously $\nu=\overline{\mu}$.
  \medskip\par
  \noindent \emph{(e)}
  \ \ It is an immediate consequence of (b) and (d).
\end{preuve}

%
\subsectionmark{Theorems \protect\ref{thm:extremal caracterisation}, \protect\ref{thm:extremal as limit} and \protect\ref{th:existence}}
\subsection{Proofs of Theorems \protect\ref{thm:extremal caracterisation}, \protect\ref{thm:extremal as limit} and \protect\ref{th:existence}}
\subsectionmark{Theorems \protect\ref{thm:extremal caracterisation}, \protect\ref{thm:extremal as limit} and \protect\ref{th:existence}}

\begin{preuve}[Proof of Theorem \ref{thm:extremal caracterisation}]
  \noindent (c)$\Rightarrow$(b)
  is immediate.
  \medskip\par
  \noindent (b)$\Rightarrow$(a)
  \ \ For $B\in\F[-\infty]$, let $\D := \bigl\{ A\in\F : \mu(A\cap B)=\mu(A)\mu(B) \bigr\}$.
  The set $\D$ satisfies\\
  $\bullet$ $\Omega\in\D$;\\
  $\bullet$ $A_1, A_2\in\D,\ A_1\subset A_2$ implies $A_2\prive A_1\in\D$, and\\
  $\bullet$ if $(A_n)_{n>0}$ is a sequence of disjoint sets of $\D$ then, $\bigcup_{n>0}A_n\in\D$.\\
  This makes $\D$ a Dynkin system and, hence, a sub-$\sigma$-algebra of $\F$.
  Moreover, by hypothesis all cylinders are in $\D$, so that $\D=\F$.
  In particular, $B\in\F$ so $\mu(B)=\mu(B)^2$ i.e. $\mu(B)\in\{0,1\}$.
  \medskip\par
  \noindent (a)$\Rightarrow$(c)
  \ \ Let $A\in\F$ and $(\Lambda_n)_{n>0}$ be an increasing sequence of time boxes which converges to $S$.
  The reverse martingale theorem yields $\mu\bigl(A\bigm|\F[\ms{(\Lambda_n)}]\bigr)\xrightarrow{L^1}\mu\big(A\mid\F[-\infty]\big)$.
  Since $\mu$ is trivial on $\F[-\infty]$, $\mu\big(A\mid\F[-\infty]\big)=\mu(A)$ $\mu$-a.s..
  We deduce that
  \[
  \forall\varepsilon>0,\quad \exists\Delta\in\Sb:\quad \mu\Bigl(\Bigl|\mu(A\mid \F[\ms{\Delta}])-\mu(A)\Bigr|\Bigr)<\varepsilon\;.
  \]
  Hence, for all $\Lambda\in\Sb : \Lambda\supset\Delta$,
  \begin{eqnarray*}
    \sup_{B\in\F[\pc{\Lambda}]} \bigl| \mu(A\cap B)-\mu(A)\mu(B) \bigr| & \leq & \sup_{B\in\F[\pc{\Delta}]} \bigl| \mu(A\cap B)-\mu(A)\mu(B) \bigr| \\
    & \leq & \sup_{B\in\F[\pc{\Delta}]} \Bigl| \mu\Bigl( \1{B}\bigl[ \mu(A\mid\F[\ms{\Delta}])-\mu(A) \bigr]\Bigr) \Bigr| \\
    & \leq & \mu\Bigl( \Bigl| \mu(A\mid\F[\ms{\Delta}])-\mu(A) \Bigr| \Bigr) \\
    & \leq & \varepsilon\;.
  \end{eqnarray*}
\end{preuve}

\begin{preuve}[Proof of Theorem \ref{thm:extremal as limit}]
  \noindent \emph{(i)}
  \ \ Since $\mu$ is consistent with $\gamma$, $\gamma_{\Lambda_n}(h)=\mu\bigl(h\bigm|\F[\pc{(\Lambda_n)}]\bigr)$.
  The reverse martingale theorem thus yields
  \[
  \gamma_{\Lambda_n}(h) \xrightarrow[\mu\text{-a.s.}]{L^1} \mu\bigl(h\bigm|\F[-\infty]\bigr) = \mu(h)
  \]
  \par
  \noindent \emph{(ii)}
  \ \ The result follows from (i) because the set of continuous local functions contains a countable dense set (for the sup-norm).
\end{preuve}

\begin{preuve}[Proof of Theorem \ref{th:existence}]
  \noindent \emph{(i)}
  If $\Lambda$ is a time box and $h$ a local continuous function,
  \begin{eqnarray*}
    \mu(h)&=&\lim_n \int \nu_n(d\omega)\,\gamma_{\Lambda_n}(h\mid \omega) \\
    & = &\lim_n \int \nu_n(d\omega)\,\gamma_{\Lambda_n}\bigl(\gamma_\Lambda(h)\bigm| \omega\bigr) \\
    &=& \mu\bigl(\gamma_\Lambda(h)\bigr)\;.
  \end{eqnarray*}
  The second identity is due to the consistency of the kernels of the POS.
  The last one follows from weak convergence and the quasilocality of $\gamma$.
  This proves consistency of $\mu$ with the POS $\gamma$.
  \medskip\par
  \noindent \emph{(ii)}
  By a slight strengthening of the Banach-Alaoglu theorem~(eg.\ Theorem 3.16 in \cite{rud73}), the space of probability measures on $\Omega$ endowed with the weak convergence is metrizable and compact.
  Hence every sequence of the form $\nu_n\gamma_{\Lambda_n}$ has a convergent subsequence whose limit is in $\G(\gamma)$ by (i).
  \medskip\par
  \noindent \emph{(iii)}
  By the argument in (ii), there is a subsequence $\Lambda_{n_i}$ and probability measures $\mu_1$ and $\mu_2$ such that $\gamma_{\Lambda_{n_i}}(\,\cdot\mid\omega_1) \to \mu_1$ and $\gamma_{\Lambda_{n_i}}(\,\cdot\mid\omega_2) \to \mu_2$ weakly.
  By (i) $\mu_1,\mu_2\in\G(\gamma)$ while the hypothesis ensures that $\mu_1\neq\mu_2$.
\end{preuve}

%
%
\section{Proofs of color-ordering inequalities}

%
\subsection{Proof of Theorem \protect\ref{thm:FKG}}

We need an auxiliary result based on the notion of coupling.
\begin{definition}
  A \dfn{coupling} $P$ between two measures $\mu$ and $\nu$ on a measurable space $X$ is a measure $P$ on $X^2$ having $\mu$ and $\nu$ as its marginals, that is such that $P(A,X)=\mu(A)$ and $P(X,A)=\nu(A)$ for all events $A$.
\end{definition}

\begin{proposition}[Strassen theorem]
  \label{prop:comparaison}
  For any two probability measures $\mu$ and $\nu$ on $\Omega$, the following statements are equivalent:
  \begin{enumerate}
  \item
    $\mu\dom\nu$
  \item
    There exists a coupling $P$ of $\mu$ and $\nu$ such that $P(\sigma\leq\sigma')=1$.
  \end{enumerate}
\end{proposition}
See \cite{Georgii-Haggstrom-Maes} for a proof.

The following theorem is the POS counterpart  of a theorem proved by Holley (see \cite{Georgii-Haggstrom-Maes}) for Gibbs fields.

\begin{theoreme}[POS-Holley]
  \label{thm:POS-Holley}
  Let $\gamma$ and $\gamma'$ be two POS on the same color space $E$, $\Lambda\in\Sb$ and $\eta,\eta'$ be two configurations on $\ms{\Lambda}$.
  If for all $x\in\Lambda$, $e\in E$ and $\xi,\xi'\in\Omega_{\Lambda\cap x_-}$ satisfying $\xi\leq\xi'$ we have
  \begin{equation}
    \label{eqn:holley}
    \gamma_x \bigl(\sigma_x\geq e\bigm|\eta \xi \bigr) \;\leq\; \gamma'_x\bigl(\sigma_x\geq e\bigm|\eta'\xi'\bigr)
  \end{equation}
  then 
  \begin{equation}
    \gamma_\Lambda(\cdot\mid\eta) \;\dom\; \gamma'_\Lambda(\cdot\mid\eta') \quad \mbox{on }\F[\pc{\Lambda}]\;.
  \end{equation}
\end{theoreme}

\begin{preuve}
  We shall construct a coupling $P_\Lambda$ between $\gamma_\Lambda(\cdot\mid\eta)$ and $\gamma'_\Lambda(\cdot\mid\eta')$ on $\F[\Lambda]$ defined by random variables $(\xi_\Lambda,\xi'_\Lambda)$ such that $P_\Lambda(\xi_\Lambda\le\xi'_\Lambda)=1$.
  By Strassen Theorem this proves stochastic dominancy on $\F[\Lambda]$.
  The extension to $\F[\pc{\Lambda}]$ follows from \reff{eq:rp5}.
  
  We start with single sites.
  For each $x\in\Lambda$ we construct a coupling $P_x$ between $\gamma_x(\cdot\mid\eta\xi)$ and  $\gamma_x(\cdot\mid\eta'\xi')$ through the random variables
  \[ \begin{split}
    \xi_x & := \max\Big\{ e\in E:\ \gamma_x \bigl(\sigma_x\geq e\bigm|\eta \xi  \bigr) \geq U_x  \Big\} \\
    \xi'_x & := \max\Big\{ e\in E:\ \gamma'_x\bigl(\sigma_x\geq e\bigm|\eta'\xi' \bigr) \geq U_x  \Big\} \\
  \end{split} \]
  where $\{U_x:x\in\Lambda\}$ is a family of independent random variables uniformly distributed on $[0,1]$.
  Denoting $P_x$ the distribution inherited by the pair $(\xi_x,\xi'_x)$ from the distribution of $U_x$, we have that, by hypothesis \reff{eqn:holley},
  \begin{equation}
    P_x(\xi_x\leq\xi'_x) \;=\;1
  \end{equation}
  while, clearly, 
  \begin{equation}\label{eq:rp6}
    P_x\big( \sigma_x\ge e\,, \sigma_x'\in E \big) \;=\; \gamma_x\bigl( \sigma_x\ge e \bigm| \eta\xi \bigr)\;.
  \end{equation}
  The reader should keep in mind that $P_x$ depends on $\xi$, $\eta$, $\xi'$ and $\eta'$, even when we are suppressing this dependency in the notation to avoid notational cluttering.

  To construct the full coupling $P_\Lambda$ we use the Reconstruction Theorem \ref{thm:reconstruction}.
  Let $y_1,\cdots,y_n$ be a sequence such that $\gamma_\Lambda=\gamma_{y_1}\cdots\gamma_{y_n}$.
  We define
  \begin{equation}
    P_\Lambda\;=\; P_{y_1}\cdots P_{y_n}
  \end{equation}
  where each $P_{y_i}$ is defined as above, and, for each realization of $\xi_{y_1},\xi'_{y_1},\cdots,\xi_{y_n},\xi'_{y_n}$, 
  \begin{eqnarray*}
    P_{y_n} &\mbox{couples}& \gamma_{y_n}(\cdot\mid\xi_{y_1\cdots y_{n-1}}\eta) \mbox{ and } \gamma_{y_n}(\cdot\mid\xi'_{y_1\cdots y_{n-1}}\eta') \\
    P_{y_{n-1}} &\mbox{couples}& \gamma_{y_{n-1}}(\cdot\mid\xi_{y_1\cdots y_{n-2}}\eta) \mbox{ and } \gamma_{y_{n-1}}(\cdot\mid\xi'_{y_1\cdots y_{n-2}}\eta') \\
    &\vdots& \\
    P_{y_1} &\mbox{couples}& \gamma_{y_1}(\cdot\mid\eta) \mbox{ and } \gamma_{y_1}(\cdot\mid\eta')\;.
  \end{eqnarray*}
  Using the inductive relation
  \begin{equation}
    P_\Lambda(A) \;=\; \int \1{A}(\xi_\Lambda,\xi'_\Lambda) \,P_{y_1\cdots y_{n-1}}(d\xi_{y_1},d\xi'_{y_1},\cdots,d\xi_{y_{n-1}},dxi'_{y_{n-1}})\, P_{y_n}(d\xi_{y_n},d\xi'_{y_n})
  \end{equation}
  it is straightforward to verify that indeed $P_\Lambda$ couples $\gamma_\Lambda(\cdot\mid\eta)$ and $\gamma'_\Lambda(\cdot\mid\eta')$, and that $P_\Lambda(\xi_\Lambda\le\xi'_\Lambda)=1$.
\end{preuve}

\begin{preuve}[Proof of Theorem \protect\ref{thm:FKG}]
  \noindent \emph{(i)}
  Proved by the previous theorem.
  \medskip\par
  \noindent \emph{(ii)}
  As a first step we show \eqref{eqn:FKG-local} for $f, g$ depending only on a single site $y$.
  Fix $\omega\in\Omega$ and denote
  \begin{equation}
    \pi(\cdot) = \gamma_y(\cdot\,g\mid\omega)\;.
  \end{equation}
  Inequality \eqref{eqn:FKG-local} is trivially true for $g=0$.
  Furthermore, for $m\in\RR$ and $\alpha>0$, the inequality $\gamma_y(fg\mid\omega)\geq\gamma_y(f\mid\omega)\gamma_y(g\mid\omega)$, implies both $\gamma_y(f\cdot(g+m)\mid\omega)\geq\gamma_y(f\mid\omega)\gamma_y(g+m\mid\omega)$ and $\gamma_y(f\cdot(\alpha g)\mid\omega)\geq\gamma_y(f\mid\omega)\gamma_y(\alpha g\mid\omega)$.
  Hence, we can suppose without loss of generality that $g$ is strictly positive and $\gamma_y(g\mid\omega)=1$ and, therefore, that $\pi$ is a probability measure on $\F[\Omega_y]$.
  If, for brevity, we denote
  \[ \begin{split}
    q (a) & = \gamma_y(\sigma_y\geq a\mid\omega) \\
    q'(a) & = \pi(\sigma_y\geq a) \\
  \end{split} \]
  we see that, by the monotonicity of $g$, $q'(a)\ge g(a)\,q(a)$ while $1-q'(a)\le g(a\bigl(1-q(a)\bigr)$.
  Hence,
  \begin{equation}\label{eq:rp12}
    \frac{q'(a)}{1-q'(a)} \;\ge\; \frac{q(a)}{1-q(a)}\;.
  \end{equation}
  Since the function $x\mapsto x/(1-x)$ is increasing, inequality \reff{eq:rp12} implies that $q(a) \leq q'(a)$, that is
  \begin{equation}
    \gamma_y\bigl(\sigma_y\geq a\bigm|\omega\bigr) \;\leq\; \pi(\sigma_y\geq a)\;.
  \end{equation}
  Thus, by the POS-Holley Theorem \ref{thm:POS-Holley}, we obtain that
  for all increasing $\F[\Omega_y]$-measurable functions $f$,
  \begin{equation}\label{eq:rp15}
    \gamma_y(f\mid\omega) \;\leq\; \pi(f)\;=\; \gamma_y(fg\mid\omega) \;=\; \frac{\gamma_y(fg\mid\omega)}{\gamma_y(g\mid\omega)}\;.
  \end{equation}
  This proves \eqref{eqn:FKG-local} for single-site increasing functions $f$ and $g$.
  
  As a second step we consider functions $f$ and $g$ that are $\F[\ms{y}]$-measurable.
  This can be reduced to the previous case through the functions $f_\omega$ and $g_\omega$ defined by $f_\omega(\eta):=f(\eta_y\,\omega_{S\prive\{y\}})$
  [This is the same trick used in the proof of Proposition \ref{prop:egalite}].
  Applying \reff{eq:rp15},
  \[ \begin{split}
    \gamma_y(fg\mid\omega) & = \gamma_y(f_\omega g_\omega\mid\omega) \\
    & \geq \gamma_y(f_\omega\mid\omega)\gamma_y(g_\omega\mid\omega) \\
    & = \gamma_y(f\mid\omega)\gamma_y(g\mid\omega) \\
  \end{split} \]
  as sought.
  
  The third and final step involves induction on the number of sites in the time box $\Delta$:
  Suppose \eqref{eqn:FKG-local} is true for all time boxes with $n$ sites and let $\Delta$ be a time box with $n+1$ sites.
  We write the kernel $\gamma_\Delta$ as $\gamma_\Delta=\gamma_{y_1}\cdots\gamma_{y_{n+1}}$ according to the Reconstruction Theorem \ref{thm:reconstruction} and denote $\Lambda:=\{y_1,\cdots,y_n\}$ and $x:=y_{n+1}$.
  Then,
  \[ \begin{split}
    \gamma_\Delta(fg\mid\omega) & = \gamma_\Lambda\bigl(\gamma_x(fg\mid\cdot)\bigm|\omega\bigr) \\
    & \geq \gamma_\Lambda\bigl(\gamma_x(f\mid\cdot)\,\gamma_x(g\mid\cdot)\bigm|\omega\bigr) \\
    & \geq \gamma_\Lambda\bigl(\gamma_x(f\mid\cdot)\bigm|\omega\bigr)\,\gamma_\Lambda\bigl(\gamma_x(g\mid\cdot)\bigm|\omega\bigr) \\
    & \geq \gamma_\Delta(f\mid\omega)\,\gamma_\Delta(g\mid\omega) \;. \\
  \end{split} \]
  The second inequality comes from the fact that $\omega\mapsto\gamma_x(f|\omega)$ is an increasing function by the POS-Holley theorem.
  
  \medskip\par
  \noindent \emph{(iii)}
  Its proof is just an application of Theorem \ref{thm:extremal as limit}.
\end{preuve}

%
\subsectionmark{{\small Theorem \protect\ref{th:color-ordering}, Proposition \protect\ref{pro:ising-stra}}}
\subsection{Proof of Theorem \protect\ref{th:color-ordering} and Proposition \protect \ref{pro:ising-stra}}
\subsectionmark{{\small Theorem \protect\ref{th:color-ordering}, Proposition \protect \ref{pro:ising-stra}}}

\begin{preuve}[Proof of Theorem \protect\ref{th:color-ordering}]
  \noindent \emph{(i)}
  Apply twice (i) of Theorem \ref{thm:FKG}.
  
  \medskip\par
  \noindent \emph{(ii)}
  Let $A:=\{\eta\in\Omega_\Delta;\ \eta_{\Delta\prive\Lambda}=\oplus_{\Delta\prive\Lambda}\}$.
  Since $\1{A}$ is an increasing function, the FKG inequality \reff{eqn:FKG-local} implies 
  \[
  \gamma_\Delta\bigl(f\1{A}\bigm|\oplus\bigr) \;\geq\; \gamma_\Delta(f\mid\oplus)\, \gamma_\Delta(A\mid\oplus)\;.
  \]
  Thus
  \[ \begin{split}
    \gamma_\Lambda(f\mid\oplus) & = \gamma_\Delta(f\mid A,\oplus) \\
    & = \frac{\gamma_\Delta\bigl(f\1{A}\bigm|\oplus\bigr)}{\gamma_\Delta(A\mid\oplus)} \\
    & \geq \gamma_\Delta(f\mid\oplus)\;. \\
  \end{split} \]
  
  \medskip\par
  \noindent \emph{(iii)}
  We construct $\mu^\oplus$ in three steps.
  First, suppose that $f$ is a local bounded increasing function and let $\Delta\in\Sb$ such that $\Supp(f)\subset\Delta$.
  Choose $\left( \Lambda_n \right)_{n\in\NN}$ an increasing sequence of time boxes such that\\
  $\bullet$ $\Lambda_0=\Delta$,\\
  $\bullet$ for all $n\in\NN$, $\ps{(\Lambda_n)}\cap\Lambda_{n+1}=\emptyset$,\\
  $\bullet$ $\lim_{n\to\infty}\Lambda_n=\Delta\cup\Delta_-$.\\
  The sequence $\left(\mu^\oplus_{\Lambda_n}(f)\right)_{n\in\NN}$ is decreasing by part (ii) and bounded because so is $f$.  
  Therefore this sequence is convergent to a limit that defines $\mu^\oplus(f)$.
  It is straightforward to see that this limit is independent of the sequence (given two such sequences there is a larger sequence satisfying the same properties and having the initial sequences as subsequences).  
  
  Second, consider a bounded local function $g$ that is not necessarily increasing.
  Such a function admits the decomposition 
  \[
  g(\omega) = \sum_{\Upsilon\subset\Supp(g)}\alpha_\Upsilon\; \1{\omega_\Upsilon=u}
  \]
  for suitable real numbers $(\alpha_\Upsilon)_{\Upsilon\subset\Supp(g)}$.
  As each $\1{\omega_\Upsilon=u}$ is an increasing function, we take advantage of the preceding definition to define
  \[
  \mu^\oplus(g) := \sum_{\Upsilon\subset\Supp(g)}
  \alpha_\Upsilon\; \mu^\oplus(\1{\omega_\Upsilon=u})
  \]
  
  Lastly, we define $\mu^\oplus(h)$ for any function $h$ through the usual limit procedure (``standard machine'' of the construction of Lebesgue integrals).
  The resulting measure inherits the limit property \reff{eq:rp20} and is consistent with $\gamma$ by (i) of Theorem \ref{th:existence}.
  
  $\mu^\ominus$ is defined analogously.
  
  \medskip\par
  \noindent \emph{(iv)--(v)}
  Both statements follow from the fact that, by (i) of Theorem \ref{thm:extremal as limit} and (i) and (iii) above,
  \begin{equation}
    \label{eq:rp21}
    \mu^\ominus \;\dom\; \mu \;\dom\; \mu^\oplus
  \end{equation} 
  for each extremal measure $\mu\in\G(\gamma)$.
\end{preuve}

\begin{preuve}[Proof of Proposition \protect \ref{pro:ising-stra}]
  \noindent
  Let $\xi,\eta$ be two configurations and $x\in S$, then
  \begin{equation}\label{eq:rp23}
    \xi \leq \eta \; \ssi\; \xi_{Nx}+\xi_{Wx} \leq \eta_{Nx}+\eta_{Wx}\;. 
  \end{equation}
  This immediately shows the validity of hypothesis \reff{eq:rfkg} for the Stavskaya model.
  For the Ising model, \reff{eq:rp23} implies  that 
  \begin{equation}
    \frac{1}{1+\exp\big( -\beta(\xi_{Nx}+\xi_{Wx}+h) \big)} \;\leq\; \frac{1}{1+\exp\big( -\beta(\eta_{Nx}+\eta_{Wx}+h) \big)} 
  \end{equation}
  which, in turns, implies $\I_x(\sigma\geq 1 \mid \xi) \;\leq\; \I_x(\sigma\geq 1 \mid| \eta)$.
\end{preuve}

%
%
\section{Proofs of the uniqueness criteria}

%
\subsection{Bounded-uniformity criterion}

\begin{preuve}[Proof of Theorem \protect\ref{thm:useless}]
  We will prove that each measure in $\G(\gamma)$ is extremal, hence $\G(\gamma)$ can not contain more than one element.
  
  Let $\mu$ be in $\G(\gamma)$ and $B\in\F[-\infty]$ such that $\mu(B)>0$.
  We will prove that $\mu(B)=1$.  To this we consider
  \begin{equation}
    \nu(\,\cdot\,)\;:\;=\;\mu(\cdot\mid B)\;=\;\1{B}(\,\cdot\,)\frac{\mu(\,\cdot\,)}{\mu(B)}\;.
  \end{equation}
  Since $\nu\ll\mu$ and $\omega\mapsto\1{B}(\omega)/\mu(B)$ is $\F[-\infty]$-measurable, $\nu\in\G(\gamma)$ [part (c) of Theorem \ref{thm:measure general prop}].
  Moreover, for all cylinders $A$,
  \begin{eqnarray*}
    \nu(A) & = & [\nu\,\gamma_\Lambda](A) \\
    & = & \mu\bigl( \nu\,\gamma_\Lambda(A) \bigr) \\
    & = & \iint\gamma_\Lambda(A,\omega)\,d\nu(\omega)\,d\mu(\xi) \\
    & \geq & \iint c\ \gamma_\Lambda(A,\xi)\,d\nu(\omega)\,d\mu(\xi) \\
    & = & c\ \nu\big( [\mu\,\gamma_\Lambda](A) \big) \\
    & = & c\ \mu(A) \\
  \end{eqnarray*}
  with $\Lambda$ chosen so to satisfy the hypotheses of the theorem.
  The fact that $\nu(A)\geq c\ \mu(A)$ for all cylinders $A$ is tantamount to $\nu\geq c\ \mu$. 
  In particular $0=\nu(\Omega\prive B)\geq c\ \mu(\Omega\prive B)$, which proves that $\mu(B)=1$.
\end{preuve}

%
\subsection{Dobrushin criterion}

We first prove two lemmas.
The first one refers to the effect of the ``broom'' $\gamma_y$ on functions that depend also on colors of sites other than $y$.

\begin{lemme}[Multisite dusting lemma]
  Let $y\in S$, $x\in \pc{y}$ and $f$ be a $\F[\pc{y}]$-measurable function.
  Then,
  \begin{equation}
    \label{eqn:mulitsite lemma}
    \delta_x\left(\gamma_yf\right)
    \begin{cases}
      = 0                                       & \text{if } x=y \\
      \leq \delta_x(f)                          & \text{if } x\in y^* \\
      \leq \delta_x(f) + \delta_y(f)\alpha_{y,x} & \text{if $x\in y_-$ (i.e. $x<y$)} \\
    \end{cases}
  \end{equation}
\end{lemme}

\begin{preuve}
  The case $x=y$ is evident.
  For the other cases, denote $f_\xi:E\mapsto\RR$ the function defined by $f_\xi(\eta_y) := f(\eta_y\xi_{S\prive\{y\}})$. 
  Let us first consider $x\in y_-$. If $\xi\pequ{x}\eta$ we have
  \[ \begin{split}
    \Bigl| \gamma_y(f\mid\xi) - \gamma_y(f\mid\eta) \Bigr| & = \Bigl| \gamma_y(f_\xi\mid\xi) - \gamma_y(f_\eta\mid\eta) \Big| \\
    & \leq \Bigl| \gamma_y(f_\xi\mid\xi) - \gamma_y(f_\eta\mid\xi) \Bigr| + \Bigl| \gamma_y(f_\eta\mid\xi) - \gamma_y(f_\eta\mid\eta) \Bigr| \\
    & \leq \gamma_y\bigl(\delta_x(f)\bigm|\xi\bigr) + \delta_x\bigl(\gamma_y(f)\bigr) \\
    & \leq \delta_x(f) + \delta_y(f)\,\alpha_{y,x} \\
  \end{split} \]
  For $x\in y^*$, the computation is the same except that the second term in the second line disappears  because $\gamma_y(f_\eta\mid\cdot)$ is $\F[y_-]$-measurable.
\end{preuve}

The second lemma establishes an order for the cleaning of sites.
\begin{lemme}
  \label{lem:suite}
  For any time-box $\Lambda$ there exists a one-to-one sequence $(x_n)_{n\geq 1}$ of sites such that
  \begin{itemize}
  \item[(1)]
    $\bigcup_{n\geq 1}\{x_n\}=\Lambda\cup\Lambda_-$
  \item[(2)]
    $\bigcup_{n\leq k}\{x_n\}\in\Sb$, for all $k\ge 1$.
  \end{itemize}
\end{lemme}
\begin{preuve}
  The first terms of the sequence are
  \[
  \bigl\{x_1,\cdots,x_{r_1}\bigr\} \;:=\; \max(\Lambda) = \max(\Lambda\cup\Lambda_-)
  \]
  This finite sequence satisfies (2) because it is a slice (see Definition \ref{def:slice} and Proposition \ref{prop:Sb}).
  The remaining terms are defined iteratively by
  \[
  \bigl\{x_{r_k+1},\cdots,x_{r_{k+1}}\bigr\} \;:=\; \max\bigl(\{x_1,\cdots,x_{r_k}\}_-\bigr) \;=\; \max\Bigl(\Lambda\cup\Lambda_-\prive\{x_1,\cdots,x_{r_k}\}\Bigr)
  \]
  This procedure exhausts $\Lambda\cup\Lambda_-$, so the sequence $(x_n)_{n\geq 1}$ satisfies (1).
  To prove (2) it is sufficient to show that if $\Delta\in\Sb$ and $x\in\max(\Delta_-)$, then $\Delta\cup\{x\}\in\Sb$.
  
  Indeed, such $x$ satisfies $x_+\cap\Delta_-=\emptyset$ and furthermore, $x\in\Delta_-$, $x_-\subset\Delta_-$ and $x_-\cap\Delta_+=\emptyset$.
  These relations imply that
  \[ \begin{split}
    \big(\Delta\cup\{x\}\big)_+ \bigcap \big(\Delta\cup\{x\}\big)_- & = \Big[\big(\Delta_+\cup x_+\big) \bigcap \big(\Delta_-\cup x_-\big)\Big] \prive\big(\Delta\cup\{x\}\big) \\
    & = \Big[\big(\Delta_-\cap x_+\big) \bigcup \big(\Delta_+\cap x_-\big)\Big] \prive\big(\Delta\cup\{x\}\big) \\
    & = \emptyset\;; \\
  \end{split} \]
  which proves that $\Delta\cup\{x\}\in\Sb$.
\end{preuve}

\begin{preuve}[Proof of Dobrushin Criterion]
  Fix a local bounded function $f$ and choose a time box $\Lambda$ such that $\Supp(f)\subset\Lambda\cup\Lambda_-$.
  Let $(y_n)_{n\in\NN^*}$ be the sequence constructed verifying Lemma \ref{lem:suite} for $\Lambda$.
  The kernel $T_n := \gamma_{y_n}\cdots\gamma_{y_2}\gamma_{y_1}$ is then well defined.
  By the multisite dusting lemma
  \[
  \Delta(T_1f) \;=\; \Delta(\gamma_{y_1}f) \;\leq\; \sum_{x\neq y_1} \Bigl[ \delta_x(f)+\delta_{y_1}(f)\,\alpha_{y_1,x} \Bigr] \;\leq\; \sum_{x\neq y_1}\delta_x(f)+\Gamma\,\delta_{y_1}(f)\;.
  \]
  By induction, for $n\geq 1$
  \[ \begin{split}
    \Delta(T_nf) & = \Delta\bigl( \gamma_{y_n}\cdots\gamma_{y_2}(\gamma_{y_1}f) \bigr) \\
    & \leq \sum_{x\neq y_2,\cdots ,y_n}\delta_x(\gamma_{y_1}f) + \Gamma\sum_{k=2}^n\delta_{y_k}\bigl(\gamma_{y_1}(f)\bigr) \\
    & \leq \sum_{x\neq y_1,\cdots,y_n}\delta_x(f) + \Gamma\sum_{k=2}^n\delta_{y_k}(f) + \delta_{y_1}(f)\Bigl(\sum_{x\neq y_1,\cdots,y_n}\alpha_{y_1,x} + \Gamma\sum_{k=2}^n\alpha_{y_1y_k}\Bigr) \\
    & \leq \sum_{x\neq y_1,\cdots,y_n}\delta_x(f) + \Gamma\sum_{k=1}^n\delta_{y_k}(f) \\
  \end{split} \]
  The last line comes from the fact that
  \[
  \sum_{x\neq y_1,\cdots,y_n}\alpha_{y_1,x} + \Gamma\sum_{k=2}^n\alpha_{y_1,y_k} \;\leq\; \sum_{x\neq y_1}\alpha_{y_1,x} \;\leq\; \Gamma
  \]
  Note that, in particular, the quasilocal function $T_nf $  has $\Delta(T_nf)<\infty$.
  
  Let $\mu, \nu\in\G(\gamma)$.
  To prove the criterion, it is sufficient to prove that for all local bounded functions $f$ we have $\mu(f)=\nu(f)$.
  By consistency, we have that for all $n\in\NN$,
  \[ \begin{split}
    \big| \nu(f)-\mu(f) \big| & = \big| \nu(T_nf)-\mu(T_nf) \big| \\
    & \leq \Delta(T_nf) \\
    & \leq \sum_{k>n}\delta_{y_k}(f)+\Gamma\sum_{k\leq n}\delta_{y_k}(f) \\
  \end{split} \]
  Letting $n$ go to infinity, we obtain 
  \begin{equation}
    \label{eq:rpp31}
    \big| \nu(f)-\mu(f) \big| \;\leq\; \Gamma\Delta(f)
  \end{equation}
  for every local bounded function $f$.
  Using approximations by local functions, this inequality extends to quasilocal functions of bounded total oscillation.
  We can, therefore, apply \reff{eq:rpp31} with $f\to T_nf$ to get
  \[ \begin{split}
    \big| \nu(f)-\mu(f) \big| & = \big| \nu(T_nf)-\mu(T_nf) \big| \\
    & \leq \Gamma\Delta(T_nf) \\
    & \leq \Gamma\big( \sum_{k>n}\delta_{y_k}(f)+\Gamma\sum_{k\leq n}\delta_{y_k}(f) \Big)\;. \\
  \end{split} \]
  Letting $n\to\infty$ we obtain $\big| \nu(f)-\mu(f) \big| \leq \Gamma^2\Delta(f)$ and, by induction, $\big| \nu(f)-\mu(f) \big| \leq \Gamma^m\Delta(f)$ for all $m\in\NN$.
  Since $\Gamma<1$, the limit $m\rightarrow\infty$ yields $\nu(f)=\mu(f)$.
\end{preuve}

\noindent \emph{Proof of Proposition \ref{prop:crm2s}.}
The proposition is a particular case of the following known fact.

\begin{proposition}
  Let $\mu$ and $\nu$ be measures on a countable space $E$ and $f$ a bounded function.
  Then,
  \begin{equation}
    \bigl|\mu(f)-\nu(f)\bigr| \;\le\; \delta(f)\,\frac12 \sum_{a\in E} \bigl|\mu(a)-\nu(a)\bigr|
  \end{equation}
  with equality if $\left|E\right|=2$.
\end{proposition}
\begin{preuve}
  We provide the proof for completeness.
  It is a particular instance of the relation between various definitions of the variational distance.
  
  A simple calculation shows that for any fixed point $e\in E$
  \begin{eqnarray}\label{eq:rpp.35}
    \mu(f)-\nu(f) &=& \sum_{a\in E} f(a)\,\mu(a) - \sum_{a\in E} f(a)\,\nu(a)\nonumber \\
    &=& \sum_{\scriptstyle a\in E\atop \scriptstyle a\neq e} \bigl[f(a)-f(e)\bigr] \bigl[\mu(a)-\nu(a)\bigr]\;.
  \end{eqnarray}
  Splitting the sum according to whether or not $a$ belongs to the set 
  \begin{equation}
    M\;=\bigl\{ a\in E : \mu(a)>\nu(a)\bigr\}
  \end{equation}  
  yields
  \begin{equation}
    \mu(f)-\nu(f) \;=\; A-B\;,
  \end{equation}    
  with 
  \begin{eqnarray*}
    A &=& \sum_{a\in M} \bigl[f(a)-f(e)\bigr] \bigl[\mu(a)-\nu(a)\bigr] \\
    B &=& \sum_{a\not\in M} \bigl[f(a)-f(e)\bigr] \bigl[\nu(a)-\mu(a)\bigr]\;.
  \end{eqnarray*}
  At this point we choose $e\in{\rm argmin} f$ so to have $A,B\ge 0$ and
  \begin{equation}
    \bigl| \mu(f)-\nu(f)\bigr| \;=\; \max(A,B)   
  \end{equation}
  [for $A$ and $B$ non-negative $\left|A-B\right|\le \max(A,B)$].
  But
  \begin{eqnarray}
    A \,,\, B &\le & \delta(f)\bigl[\mu(M)-\nu(M)\bigr]\nonumber \\
    &=& \delta(f)\bigl[\nu(M^{\rm c})-\mu(M^{\rm c})\bigr] \\
    &=& \delta(f)\,\frac12 \Bigl\{\bigl[\mu(M)-\nu(M)\bigr] + \bigl[\nu(M^{\rm c})-\mu(M^{\rm c})\bigr]\Bigr\} \\
    &=& \delta(f)\,\frac12 \sum_{a\in E} \bigl|\mu(a)-\nu(a)\bigr|\;.
  \end{eqnarray}
  
  If $E=\{a,e\}$ \reff{eq:rpp.35} becomes
  \begin{equation}
    \bigl| \mu(f)-\nu(f)\bigr| \;=\; \bigl[f(a)-f(e)\bigr] \bigl[\mu(a)-\nu(a)\bigr]\;,
  \end{equation} 
  hence $\bigl| \mu(f)-\nu(f)\bigr| = \delta(f) \bigl|\mu(a)-\nu(a)\bigr|$.
\end{preuve}

%
\subsectionmark{disagreement percolation}
\subsection{Oriented disagreement percolation criterion}
\subsectionmark{disagreement percolation}

Let us start by introducing some standard notation.
\begin{definition}
  Let $\Gamma\subset S$, $\mu$ and $\nu$ measures on $(\Omega,\F)$ and $\Delta\subset\Gamma$.
  The \dfn{variational distance} of $\mu$ and $\nu$ (projected) on $\Delta$ is
  \begin{equation}
    \label{eqn:var norm:sup-sum}
    \|\mu-\nu\|_\Delta \;\bydef\; \frac{1}{2} \sum_{\omega\in\Omega_\Delta} \bigl|\mu(\omega)-\nu(\omega)\bigr|\;.
  \end{equation} 
\end{definition}
[A well known argument shows that $\|\mu-\nu\|_\Delta := \sup_{A\in\F[\Delta]}|\mu(A)-\nu(A)|$, which is the expression used to define variational distances on non-countable spaces.
Expression \reff{eqn:var norm:sup-sum} is more useful in the countable setting supposed here.]

In the proof that follows we shall exploit the identity
\begin{equation}
  \|\mu-\nu\|_\Delta \;=\; \min\Bigl\{Q_\Delta\bigl(\{\sigma_\Delta\neq\sigma'_\Delta\}\bigr) : Q_\Delta \mbox{ coupling of } \mu_\Delta \mbox{ and }\nu_\Delta\Bigr\}
\end{equation}
where $\mu_\Delta$ and $\nu_\Delta$ are the projections of the measures $\mu$ and $\nu$ to $\F[\Delta]$.
This remarkable equality (see e.g. pp.~61--62 in~\cite{dob96a} for a simple proof) conveys two pieces of information.
First it relates the variational distance with the \emph{Kantorovich-Wasserstein distance} defined by the right-hand side.
Second, it states that there exists a coupling that realizes the equality.
As a matter of fact, this coupling can be defined in a relatively simple explicit way [formula (14.33) in~\cite{dob96a} or Chapter 3 in~\cite{Ferrari-Galves}), though in the sequel only its existence plays a role.

\begin{definition}
  An \dfn{optimal $\Delta$-coupling} for the measures $\mu$ and $\nu$ is a measure $P_\Delta$ on $(\Omega^2,\F[\Delta]^2)$ such that
  \begin{equation}
    \|\mu-\nu\|_\Delta \;=\; P_\Delta\bigl(\{\sigma_\Delta\neq\sigma'_\Delta\}\bigr) \;.
  \end{equation}
\end{definition}
Such a coupling translates disagreement into distance between measures.

\begin{definition}
  For $x<y\in S$ let
  \begin{equation}
    (y\join[\neq] x)\;=\; \Bigl\{(\sigma,\sigma')\in\Omega^2:\ \exists (y_k)_{1\leq k\leq n}
    \mbox{ with }y_1=y,\ y_n=x,\ y_k\in\db y_{k+1},\ \sigma_{y_k}\neq \sigma'_{y_k}\Bigr\}\;.
  \end{equation}
  This is the event ``there exists a downward-oriented path of disagreement from $y$ to $x$''.
\end{definition}

In the sequel the notation $(\Delta\join[>]\Lambda)$, for $\Delta,\Lambda\subset S$, stands for $\big\{\exists x\in\Delta,\ \exists y\in\Lambda:\ (x\join[>] y)\big\}$.
Likewise for $(\Delta\join[\neq]\Lambda)$.
Furthermore we shall denote $\psi_{\textbf{p},\Lambda}$ the restriction (projection) or $\psi_{\textbf{p}}$ to $\F[\Lambda]$.
The following proposition is the key tool in the proof of the criterion.

\begin{proposition}
  Let $\gamma$ be a POMM, $\Lambda$ a time box and $\eta$, $\eta'$ two configurations.
  Then, there exists an optimal $\Lambda$-coupling $P_\Lambda=P_{\Lambda,\eta,\eta'}$ of $\gamma_\Lambda(\cdot\mid\eta)$ and $\gamma_\Lambda(\cdot\mid\eta')$ such that:
  \begin{enumerate}
  \item[(i)]
    $\forall x\in\Lambda,\ \{\sigma_x\neq \sigma'_x\} = (\db\Lambda\join[\neq]x)\ P_\Lambda$-a.s.,
  \item[(ii)]
    the law of $\left(\1{\{\sigma_x\neq \sigma'_x\}}\right)_{x\in\Lambda}$, denoted by $P_\Lambda^{\neq}$, is such that $P_\Lambda^{\neq}\dom\psi_{\textbf{p}^\gamma,\Lambda}$.
  \end{enumerate}
\end{proposition}

\begin{preuve}
  The coupling is constructed iteratively on sets $\ms\Delta$ with $\Delta\subset\Lambda$ decreasing from $\Lambda$ to the empty set.
  The algorithm is as follows.
  \medskip\par
  \noindent \emph{Initial step.}
  Set $\Delta=\Lambda$, and define $(\sigma_{\ms{\Lambda}}, \sigma'_{\ms{\Lambda}})=(\eta_{\ms{\Lambda}}, \eta'_{\ms{\Lambda}})$.
  \medskip\par
  \noindent \emph{Iteration step.}
  Suppose that $(\sigma,\sigma')$ has already been defined on $\ms\Delta$ for a non-empty set $\Delta\subset\Lambda$ and is realized as a pair $(\sigma_{\ms{\Delta}}, \sigma'_{\ms{\Delta}})$ with $(\sigma_{\ms{\Lambda}}, \sigma'_{\ms{\Lambda}})=(\eta_{\ms{\Lambda}}, \eta'_{\ms{\Lambda}})$.
  Pick $x\in\min\Delta$ such that there exists some $y\in\db x\subset\Delta_-$ satisfying $\xi_y\neq\xi'_y$.
  If such an $x$ does not exist, then $\gamma_\Delta(\,\cdot\mid\xi)=\gamma_\Delta(\,\cdot\mid\xi')$ on $\F[\Delta]$ and we define $\sigma_\Delta=\sigma'_\Delta$ (obviously, an optimal coupling). 
  If such an $x$ exists, we choose $(\sigma_x,\sigma'_x)$ distributed according to an optimal coupling $P_x$ of the single-site distributions $\gamma_x(\cdot\mid\sigma)$ and $\gamma_x(\cdot\mid\sigma')$ restricted to $\F[x]$.
  This defines a coupling $(\sigma_{\{x\}\cup\ms{\Delta}},\sigma'_{\{x\}\cup\ms{\Delta}})$.
  Notice that, restricted to $(\Lambda\prive\Delta)\cup\{x\}$ the coupling law satisfies 
  \begin{equation}\label{eq:rpp40}
    P_{(\Lambda\prive\Delta)\cup\{x\}} \;=\; P_x\,P_{\Lambda\prive\Delta}\;.
  \end{equation}
  We repeat the procedure replacing $\Delta$ by $\Delta\prive\{x\}$.
  \medskip
  
  It is clear that the algorithm above stops after finitely many iterations when $\Delta$ becomes the empty set.
  The fact that our construction defines a coupling of $\gamma_\Lambda(\cdot,\eta)$ and $\gamma_\Lambda(\cdot,\eta')$ follows inductively from \reff{eq:rpp40} and the Reconstruction Theorem \ref{thm:reconstruction}.
  Property (i) is evident from the construction, since disagreement at a site is only possible if a path of disagreement leads from this site to the boundary $\db\Lambda$.
  Regarding (ii), we see that, since at each site $x\in\Lambda$ we have chosen an optimal coupling, the iteration relation \reff{eq:rpp40} shows that the complete $\Lambda$-coupling is also optimal.
  Furthermore, if $x\in\Lambda$, $\eta,\eta'\in\Omega$ and $\xi,\xi'\in\Omega_{\Lambda\cap\ms{x}}$.
  \[ \begin{split}
    P_\Lambda\Bigl( \sigma_x\neq \sigma'_x\ \Bigm| \ (\sigma,\sigma')=(\xi\eta,\xi'\eta')\text{ on }\ms{x}\cup\ms{\Lambda} \Bigr) & = P_x\Bigl( \sigma_x\neq \sigma'_x\ \Bigm| \ (\sigma,\sigma')=(\xi\eta,\xi'\eta')\text{ on }\ms{x}\cup\ms{\Lambda} \Bigr) \\
    & = \big\| \gamma_x(\,\cdot\mid\xi\eta)-\gamma_x(\,\cdot\mid\xi'\eta') \big\|_x \\
    & \leq p_x^\gamma\;. \\
  \end{split} \]
  The optimality of $P_x$ explains the second line.
  Therefore, the POS-Holley Theorem~\ref{thm:POS-Holley} applied to the partially oriented kernels defined by $P_\Lambda^{\neq}$ and $\psi_{\textbf{p}^\gamma,\Lambda}$ shows that $P_\Lambda^{\neq} \dom \psi_{\textbf{p}^\gamma,\Lambda}$.
\end{preuve}

\begin{preuve}[Proof of Theorem \ref{thm:dp criterion}]
  We use the coupling created in the last proposition.
  Let $\mu,\nu\in\G(\gamma)$, and $\Delta,\Lambda$ be two time boxes such that $\Delta\subset\Lambda$.
  We have
  \[ \begin{split}
    \|\mu-\nu\|_\Delta & \leq \sup_{\eta,\eta'\in\Omega_{\ms{\Lambda}}} \left\|\gamma_\Delta(\,\cdot\mid\eta)-\gamma_\Delta(\,\cdot\mid\eta')\right\|_\Delta \\
    & \leq \sup_{\eta,\eta'\in\Omega} \left\|\gamma_\Lambda(\,\cdot\mid\eta)-\gamma_\Lambda(\,\cdot\mid\eta')\right\|_\Delta \\
    & = \sup_{\eta,\eta'\in\Omega} P_{\Lambda,\eta,\eta'}\left(\sigma\neq \sigma'\ in\ \Delta\right) \\
    & \leq \sup_{\eta,\eta'\in\Omega} P_{\Lambda,\eta,\eta'}\left(\exists x\in\Delta,\ \sigma_x\neq \sigma'_x\right) \\
    & \leq \sup_{\eta,\eta'\in\Omega} P_{\Lambda,\eta,\eta'}\left(\Delta\join[\neq]\db\Lambda\right) \\
    & \leq \psi_{\textbf{p}^\gamma,\Lambda}\left(\Delta\join[>]\db\Lambda\right)\;. \\
  \end{split} \]
  The third line is due to the optimality of the coupling and the last one to (ii) or the previous proposition.
  
  By letting $\Lambda$ tend to $S$, we get
  \[
  \|\mu-\nu\|_\Delta \leq \psi_{\textbf{p}^\gamma}\left(\Delta\join[>]-\infty\right)\;.
  \]
  The right-hand side is zero if $\psi_{\textbf{p}^\gamma}$ does not percolate, thus $\mu$ and $\nu$ coincide on all time boxes.
\end{preuve}

\begin{preuve}[Proof of Corollary \ref{coro:dp criterion}]
  By hypothesis there exists $q$ such that $\sup_{x\in S}p_x^\gamma<q<p_c^+(S)$.
  Therefore, by the POS-Holley Theorem~\ref{thm:POS-Holley} $\psi_{\textbf{p}^\gamma}\dom\psi_q$ and, as a consequence, $\psi_{\textbf{p}^\gamma}$ does not percolate. 
\end{preuve}

\section*{Acknowledgments} 
It is a pleasure to thank Elise Janvresse, Christof K\"ulske, Thierry de la Rue and Yvan Velenik for very useful discussions and clarifications. 

%
%
\addcontentsline{toc}{part}{Bibliography}
\bibliographystyle{alpha}
\bibliography{biblio}

\end{document}